\documentclass{aastex63}
\usepackage{graphicx}
\DeclareGraphicsExtensions{.png,.pdf,.eps,.ps}
\usepackage{url}
\usepackage{hyperref}
\hypersetup{colorlinks=false,pdfborder={0 0 0}}
\usepackage{subfigure} 
\usepackage[utf8]{inputenc}
\usepackage[english]{babel}
\usepackage{booktabs} 
\usepackage{color}
\usepackage[normalem]{ulem}

\shortauthors{Cunningham et al.}
\shorttitle{Young EB Triple System in Stellar String Theia~301}

\def\kms{km~s$^{-1}$}
\def\msun{M$_\odot$}
\def\rsun{R$_\odot$}

\def\teff{$T_{\rm eff}$}
\def\logg{$\log g$}

\begin{document}

\title{A KELT-\textit{TESS} Eclipsing Binary in a Young, 
Triple System Associated with the Local ``Stellar String" Theia~301}

\author[0000-0002-7110-3475]{Joni-Marie C.\ Cunningham}
\affil{Vanderbilt University, Department of Physics \& Astronomy, 6301 Stevenson Center Ln., Nashville, TN 37235, USA}

\author[0000-0002-2457-7889]{Dax L.\ Feliz}
\affil{Vanderbilt University, Department of Physics \& Astronomy, 6301 Stevenson Center Ln., Nashville, TN 37235, USA}

\author{Don M. \ Dixon}
\affil{Vanderbilt University, Department of Physics \& Astronomy, 6301 Stevenson Center Ln., Nashville, TN 37235, USA}

\author[0000-0002-3827-8417]{Joshua Pepper}
\affil{Department of Physics, Lehigh University, Bethlehem, PA 18015, USA}

\author[0000-0002-3481-9052]{Keivan G. Stassun}
\affil{Vanderbilt University, Department of Physics \& Astronomy, 6301 Stevenson Center Ln., Nashville, TN 37235, USA}

\author[0000-0001-5016-3359]{Robert J.\ Siverd}
\affil{Vanderbilt University, Department of Physics \& Astronomy, 6301 Stevenson Center Ln., Nashville, TN 37235, USA}

\author[0000-0002-4891-3517]{George Zhou}
\affil{Observatoire de Gen\`{e}ve}

\author[0000-0001-6023-1335]{Daniel Bayliss}
\affil{Observatoire de Gen\`{e}ve}

\author[0000-0001-5603-6895]{Thiam-Guan Tan}
\affil{Perth Exoplanet Survey Telescope}

\author[0000-0002-1617-8917]{Phillip Cargile}
\affil{Vanderbilt University, Department of Physics \& Astronomy, 6301 Stevenson Center Ln., Nashville, TN 37235, USA}

\author[0000-0001-5160-4486]{David James}
\affil{Cerro Tololo Inter-American Observatory, Colina El Pino, S/N, La Serena, Chile}

\author[0000-0002-4236-9020]{Rudolf B.\ Kuhn}
\affil{South African Astronomical Observatory, P.O. Box 9, Observatory 7935, Cape Town, South Africa}
\affil{Southern African Large Telescope, P.O. Box 9, Observatory 7935, Cape Town, South Africa}

\author[0000-0002-5365-1267]{Marina Kounkel}
\affil{Western Washington University}

\begin{abstract} 

HD~54236 is a nearby, wide common-proper-motion visual pair that has been previously identified as likely being very young by virtue of strong X-ray emission and lithium absorption. Here we report the discovery that the brighter member of the wide pair, HD~54236A, is itself an eclipsing binary (EB), comprising two near-equal solar-mass stars on a 2.4-d orbit. It represents a potentially valuable opportunity to expand the number of benchmark-grade EBs at young stellar ages. Using new observations of \ion{Ca}{2}~H\&K emission and lithium absorption in the wide K-dwarf companion, HD~54236B, we obtain a robust age estimate of $225 \pm 50$~Myr for the system. This age estimate and \textit{Gaia} proper motions show HD~54236 is associated with Theia~301, a newly discovered local ``stellar string", which itself may be related to the AB~Dor moving group through shared stellar members.
Applying this age estimate to AB~Dor itself alleviates reported tension between observation and theory that arises for the luminosity of 
AB~Dor~C when younger age estimates are used. 

\end{abstract}


\section{Introduction}\label{intro}

Eclipsing binary (EB) stars are foundational to stellar astrophysics by virtue of providing all of the fundamental physical properties of stars to high accuracy, with which to test and benchmark theoretical stellar evolution models, and to probe the ages of interesting Galactic populations. In a review of the field, \citet{Torres:2012} identified approximately 100 EBs whose parameters have been determined with sufficient accuracy to serve as stringent model benchmarks. Only one of these benchmark-grade EBs includes stars still in the pre--main-sequence (PMS) phase of evolution \citep[V1174 Ori;][]{Stassun:2004}. In addition, only about 25 of these EBs have measured metallicities, and an even smaller number have independent age constraints (as provided by, e.g., membership in clusters or moving groups), limiting the ability of even these benchmark-grade EBs to serve as the most stringent tests of stellar models. Additional benchmarks for early stellar evolution are critical in building our understanding of stellar formation. 

A number of ground and space based exoplanet transit surveys are operating that find EBs in large numbers, either as unintentional false positives or as intentional targets. The KELT survey \citep{Pepper:2007, Pepper:2012}, has so far identified four bright transiting exoplanets \citep{Siverd:2012, Beatty:2012, Pepper:2012, Collins:2013}, in conjunction with a very large number of EBs \citep[e.g.][]{Pepper:2008}.  

In the course of commissioning the KELT-South telescope we identified the bright, X-ray source HD~54236A as an EB.  It is listed in the SIMBAD database as a PMS star and listed in the SACY catalog as possessing Li in its spectrum, a diagnostic of stellar youth. 
It also possesses a wide common-proper-motion (CPM) companion, HD~54236B, which in turn provides the opportunity for independent checks on key parameters such as system age.
Most recently, the Transiting Exoplanet Survey Satellite (\textit{TESS}) observed the HD~54236 system \citep[HD~54236A has TIC ID 238162238 in the TESS Input Catalog;][]{Stassun:2019}.  

At the same time, the advent of {\it Gaia\/} has revolutionized our understanding of stellar multiples \citep[e.g.,][]{Oelkers:2018}, stellar associations \citep[e.g.,][]{Oh:2017}, and the 3D structure of the solar neighborhood \citep[e.g.,][]{Kounkel:2019}. Indeed, very recent investigations of the full phase-space structure of stars in the solar neighborhood reveal previously unrecognized ``stellar strings", long filamentary groups of young stars that appear to encode the large spiral-arm structures from which they formed \citep{Kounkel:2020}. Thus, HD~54236 represents a valuable opportunity to expand the number of known benchmark-grade EBs at young stellar ages, and to test recent suggestions of young ``stellar strings" in the solar neighborhood. 

In \S\ref{sec:star} we describe the overall context of the HD~54236 system, including the presence of HD~54236B as a known, wide CPM tertiary. In \S\ref{sec:data} we present the photometric observations and the spectroscopic observations obtained for the system used to identify HD~54236A as a previously unknown EB, and to determine the basic physical properties of the stellar components. 
In \S\ref{sec:results} we present the results of our analysis, including orbital solutions, light curve extracted stellar parameters including temperature and radius estimates for both components in the EB, SED modeling and resultant measurement of the stellar masses, and determination of the system age through consideration of the lithium abundances and the \ion{Ca}{2}~H\&K activity. Finally, in \S\ref{sec:discussion} we consider 
the membership of HD~54236 in relation to a newly discovered ``stellar string", Theia~301, which itself is related to the AB~Dor moving group through common membership between their filamentary stars. We conclude in \S\ref{sec:summary} with a brief summary of our findings.

\section{The HD~54236 System}\label{sec:star}

HD~54236A is a $V$=9.26 object in the constellation Puppis. It was identified as a member of a common proper motion binary in the first CCDM catalog \citep{Soderblom:1993}. It is also listed as a visual binary in the Washington Visual Double Stars catalog \citep{Mason:2001}. It was identified as a possible pre-main sequence (PMS) object from the SACY survey \citep{Torres:2006}: first through its detection as a strong X-ray source, and then through Li observed in its spectrum. \citet{Torres:2006} classified HD~54236A as spectral type G0V, and HD~54236B as spectral type K7V.

The total proper motion is $17.25 \pm 0.22$ mas yr$^{-1}$ for both HD~54236 A and B \citep[values from the Gaia archive][]{GaiaCollaboration:2018}, and the equivalent width (EW) of the Li line is given in the SACY catalog as 120 m\AA\ for A and 40 m\AA\ for B. No uncertainties are provided in the SACY catalog for the Li EW values. Below we report updated precise measurements of the Li EW and abundances for the stars. 

\begin{figure}[!ht]
\begin{center}
\includegraphics[width=\linewidth]{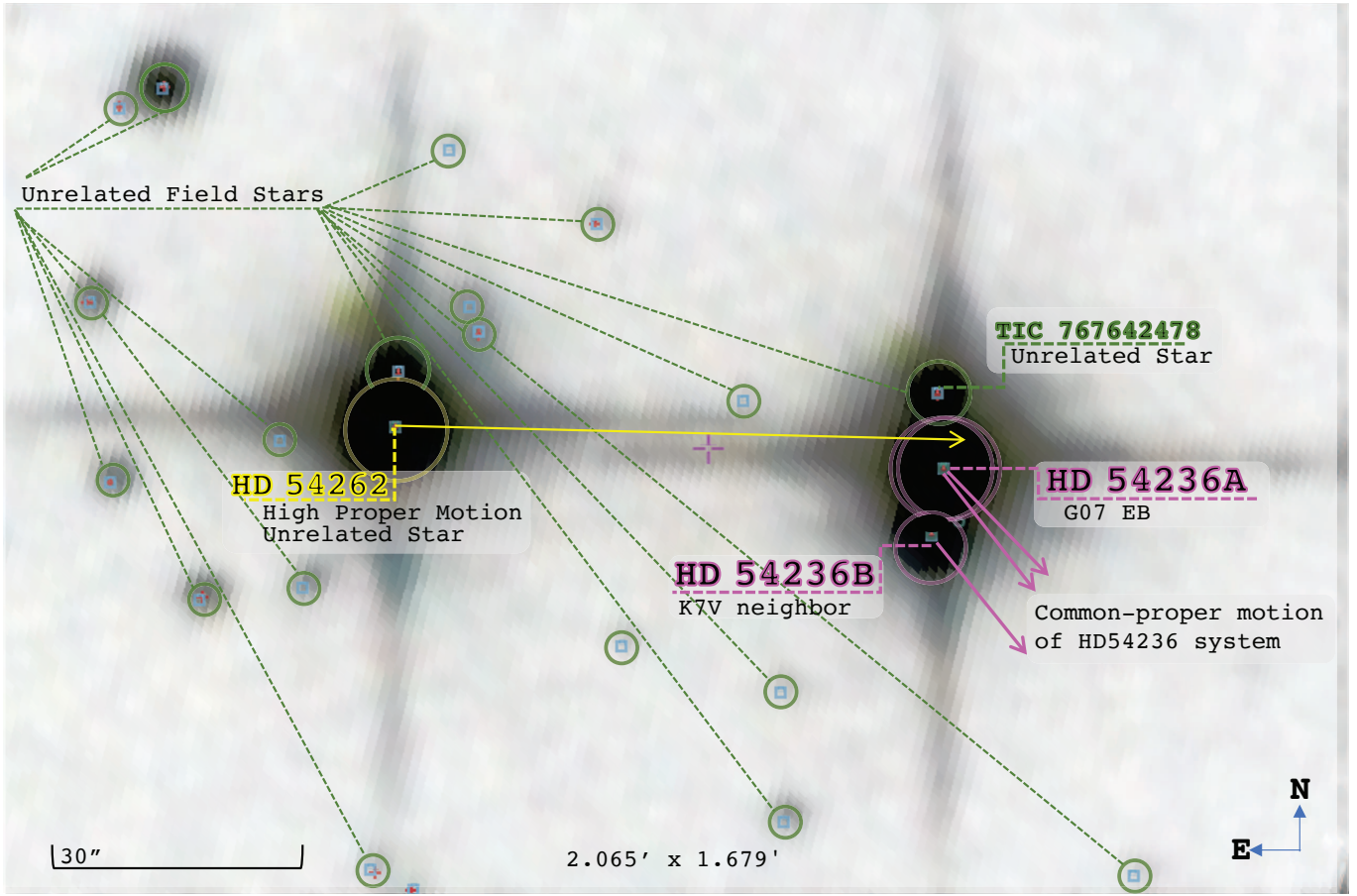} 
\caption{DSS image of the HD~54236 tertiary system and its common-proper-motion is highlighted in fushcia. The tertiary system includes HD~54236A (the eclipsing binary), and HD~54236B, the CPM tertiary member $\sim$6.5 arcseconds south of HD~54236A. Surrounding, unrelated stars (like TIC~767642478, the distant faint star to the north) are shown in green. HD~54262, an unrelated high proper motion star to the east is highlighted in yellow.}
\label{fig:finder}
\end{center}
\end{figure}

HD~54236A is in a fairly crowded part of the sky, and thus all our photometry to date is blended with nearby stars at some level.  We will pay special attention in \S~\ref{sec:photo_obs} to describing the degree of blending in each set of observations.  To clarify the environment, Figure~\ref{fig:finder} displays a DSS image of the HD~54236 system, the eclipsing binary HD~54236A and its common proper motion (CPM) companion, HD~54236B at a distance of $\sim$ 21 AU. Surrounding, unrelated field stars are also identified. This observational ecosystem includes:

\begin{itemize}
\item HD~54236A ($V$=9.26), spectral type $\sim$G0, which is the principal object of study in this paper and which we identify as an eclipsing binary. 
\item HD~54236B, which is 6.5 arcseconds south of HD~54236A and is $V$=13.2, spectral type $\sim$K7. 
It is a common proper-motion companion of HD~54236A, and we confirm below that it shares the systemic radial velocity of HD~52236A also. 
\item The bright star HD~54262, which is 54 arcseconds east of HD~54236A and is $V$=9.31. 
Based on its reported proper motion of $\mu_{\alpha} = -55.06 \pm 0.06$ mas~yr$^{-1}$ and $\mu_{\delta} = -1.36 \pm 0.06$ mas~yr$^{-1}$, compared to the proper motion of the HD~54236 system of $\mu_{\alpha} = -8.01 \pm 0.06$ mas~yr$^{-1}$ and $\mu_{\delta} = -15.28 \pm 0.05$ mas~yr$^{-1}$ (\textit{Gaia}), it appears certain that HD~54262 is not associated with the HD~54236 system. 
\item A very distant \citep[$\sim$3529~pc;][]{GaiaCollaboration:2018} and faint star ($V=14.6$) that is 7.4 arcseconds to the north of HD~54236A (TIC~767642478) is not physically associated with the HD~54236 system. 
\end{itemize}


Thus, to our knowledge, the HD~54236 system physically comprises three stars: HD~54236A which we identify below as an eclipsing binary (two stars, unresolved) and HD~54236B as a wide CPM tertiary companion that is visually resolved with a separation from HD~54236A of 6.5 arcseconds.

\section{Data}\label{sec:data}

\subsection{Photometric Observations}\label{sec:photo_obs}

\subsubsection{KELT-South Photometry}\label{sub:kelt}
The Kilodegree Extremely Little Telescope-South (KELT-South) is a dedicated exoplanet transit telescope located at the Sutherland observing station of the South African Astronomical Observatory (SAAO). The telescope, hardware, and the primary exoplanet survey is described in \citet{Pepper:2012}. 
The KELT observations are taken with a Kodak Wratten \#8 red-pass filter, resembling a widened 
Johnson-Cousins $R$-band.

KELT-South conducted a commissioning run from Jan 4, 2010, to Feb 19 2010.  During that run, KELT-South continuously observed a field centered at $\alpha$ = 08:16:00, $\delta$ = -54:00:00, with repeated 30~s exposures.  After removal of poor-quality images, we have a total of 3,041 images across 29 individual nights.  We generated relative photometry from flat-fielded images using a modified version of the \textrm{ISIS} image subtraction package \citep[see also][]{Alard:1998,Alard:2000,Hartman:2004}, in combination with point-spread function fitting using the stand-alone \textrm{DAOPHOT II} \citep{Stetson:1987, Stetson:1990} package, and the \textrm{SExtractor} program \citep{Bertin:1996}.  A more complete explanation of our pipeline procedures, and the algorithms, are 
provided in \citet{Siverd:2012}.

\begin{figure}[!ht]
\begin{center}
\includegraphics[scale=0.7,trim=20 0 0 30,clip]{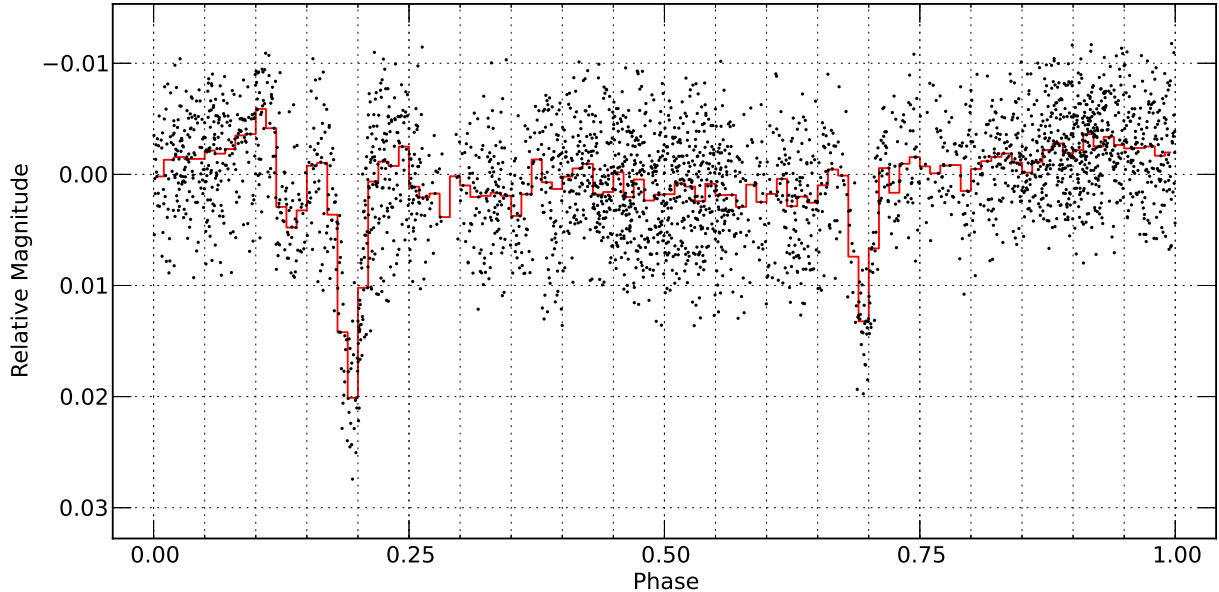} 
\caption{Phase-folded lightcurve of HD~54236A from KELT. Small symbols represent individual measurements, red lines represent the binned light curve.}
\label{fig:keltraw}
\end{center}
\end{figure}

One of the transit candidates was HD~54236A. Although the depth and out-of-eclipse variation of the lightcurve indicated that the target was unlikely to be a transiting planet, the prospect that it represented a bright PMS eclipsing binary prompted us to gather additional data of this target.  The KELT light curve of HD~54236A is shown in Fig.~\ref{fig:keltraw}. 

Because of the very wide-field nature of the KELT observing setup, with the accompanying large pixel scale (23 arcseconds per pixel), the KELT lightcurve of HD~54236A (Fig.~\ref{fig:keltraw}) is blended with all of the other stars identified in Figure \ref{fig:finder}.  The largest component of blended flux comes from HD 54262, which is about 0.33 mag fainter than HD~54236A.  Therefore, the eclipse depths seen by KELT are at least 40\% shallower than the true depths.  

An initial periodicity analysis yielded a most likely orbital period of $\sim$2.4~d. In addition, there is an out-of-eclipse variation of a nearly sinusoidal morphology in the KELT light curve with a best-fit period of 2.3955~d, i.e., slightly shorter period than the orbital (eclipse) period. A more detailed examination of the out-of-eclipse variations is provided by the \textit{TESS} light curve (see below). 

\subsubsection{PEST Photometry}\label{sub:pest}
One of the partners of the KELT-South survey is the Perth Exoplanet Survey Telescope (PEST), operated by T.G.\ Tan.  PEST consists of a 12" Meade LX200 SCT f/10 telescope, with a focal reducer yielding f/5. The camera is an SBIG ST-8XME, with 1.2 arcsec/pixel.  

PEST observed HD~54236A on the night of April 25, 2012, for 119 minutes around the predicted time of primary transit in the Cousins $R$-band, although the ephemeris was poor given the elapsed time between the original observations and the PEST observations (over 2 years).  The observations caught the last hour of the primary eclipse egress plus an hour after the eclipse.  This light curve is more pure than the discovery KELT light curve in that it excludes light contamination from HD~54262. The PEST light curve does however still include light contamination from the faint visual companion HD~54236B.  The raw photometry of the combined HD~54236A/B system was extracted using the \textrm{C-Munipack} software package\footnote{http://c-munipack.sourceforge.net/} (written by David Motl) to perform aperture photometry.  Relative photometry was derived using a set of three nearby comparison stars via a custom-written program.

\begin{figure}[!ht]
\begin{center}
\includegraphics[scale=0.5,trim=0 0 0 25,clip]{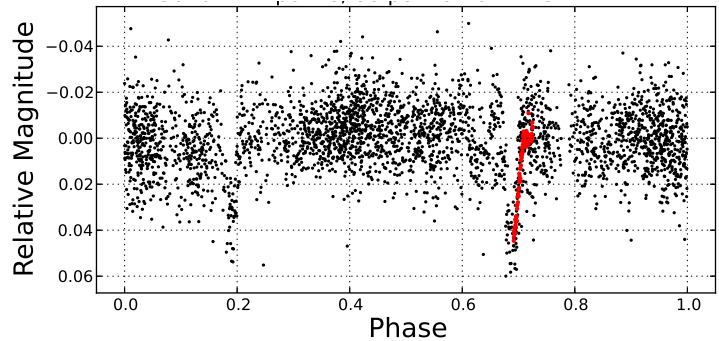} 
\caption{Phase-folded lightcurve of HD~54236A from KELT ({\it black points}) with the follow-up data from PEST over-plotted ({\it red points}).  Since the KELT data includes the flux from the unrelated high proper motion star HD~54262, the KELT data are multiplied by a factor of 3 to match the egress slope from the PEST data.}\label{fig:lc}
\end{center}
\end{figure}

The PEST light curve is shown in Figure \ref{fig:lc}, together with the KELT data scaled by a factor of 3 in flux to account for the additional dilution of the KELT data by the high proper motion bright star HD~54262. The addition of the PEST photometry allowed us to refine the orbital period from the KELT data because of the expanded time baseline.

\subsubsection{\textit{TESS} Photometry}\label{sub:tess}
\begin{figure}[!ht]
\begin{center}
\includegraphics[width=0.7\textwidth,trim=0 0 0 30,clip]{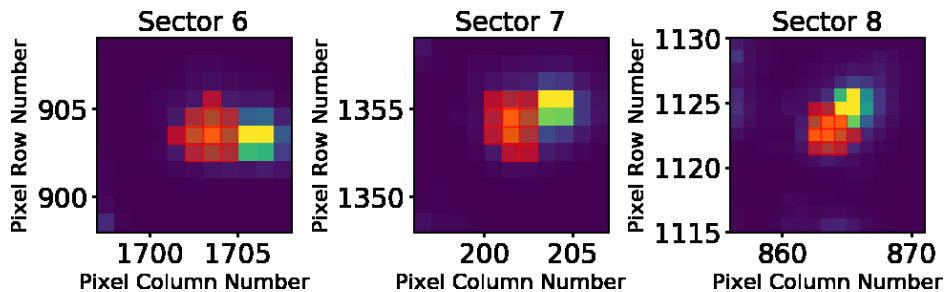}
\caption{\textit{TESS} Target Pixel File cutouts of HD~54236 for Sectors 6, 7, and 8. The red aperture mask is the optimal aperture from the SPOC pipeline, chosen to avoid the distant faint star to the north, TIC~767642477, from blending the photometry.}
\label{fig:tess_tpfs}
\end{center}
\end{figure}

\textit{TESS} observed HD~54236 (TIC~238162238) in Sectors 6, 7, and 8 from December 15th, 2018, to February 27th, 2019. The 2-minute cadence light curves were processed using the \textrm{Lightkurve} Python module \citep{Lightkurve:2018}. We extracted the light curves using the optimized aperture selected by the Science Processing Operations Center (SPOC) pipeline. The mask was chosen to avoid the high proper motion star to the west, HD~54262, from blending the photometry of HD~54236 (see Figure~\ref{fig:tess_tpfs}). 
The tertiary star, HD~54236B, remains blended but as it contributes only 2.5\% of the flux we do not attempt to correct for it. 

To estimate the background, we required a brightness threshold such that the pixels in the Target Pixel File that are 0.1\% times the standard deviation below the overall median flux of pixels in the selected aperture. We then subtracted the background from our source signal and normalized by the median flux.

\begin{figure}[!ht]
\begin{center}
\includegraphics[width=0.95\textwidth,trim=0 0 0 25,clip]{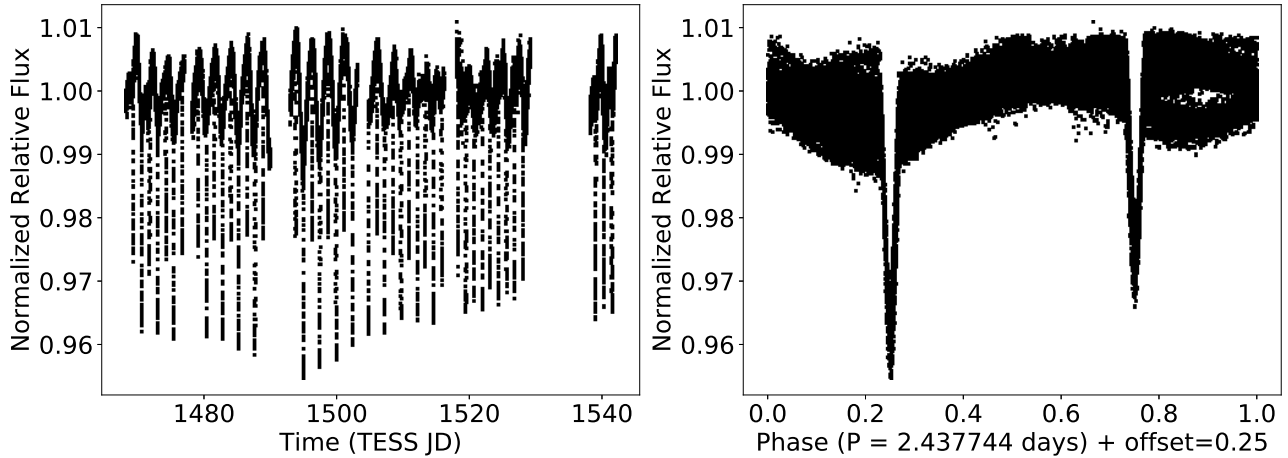} 
\caption{\textit{Left:} \textit{TESS} light curve of HD~54236 in Sectors 6, 7, and 8. \textit{Right:} Phase-folded \textit{TESS} light curve.}
\label{fig:tess_lc}
\end{center}
\end{figure}

To remove long-term trends and to assist with normalization of the flux baselines between sectors and satellite downlink times, we utilized a Biweighted Midcorrelation filter (via the \textrm{Wotan} package; \citealt{Hippke:2019}). We chose a window size of 3 times the $\sim$2.4~d period to ensure at least 3 transits are included in each window and will not be overfitted by the smoothing function. The long-term detrended light curve is shown in Figure~\ref{fig:tess_lc}.

\begin{figure}[!ht]
    \centering
    \includegraphics[scale=0.55,trim=0 0 0 25,clip]{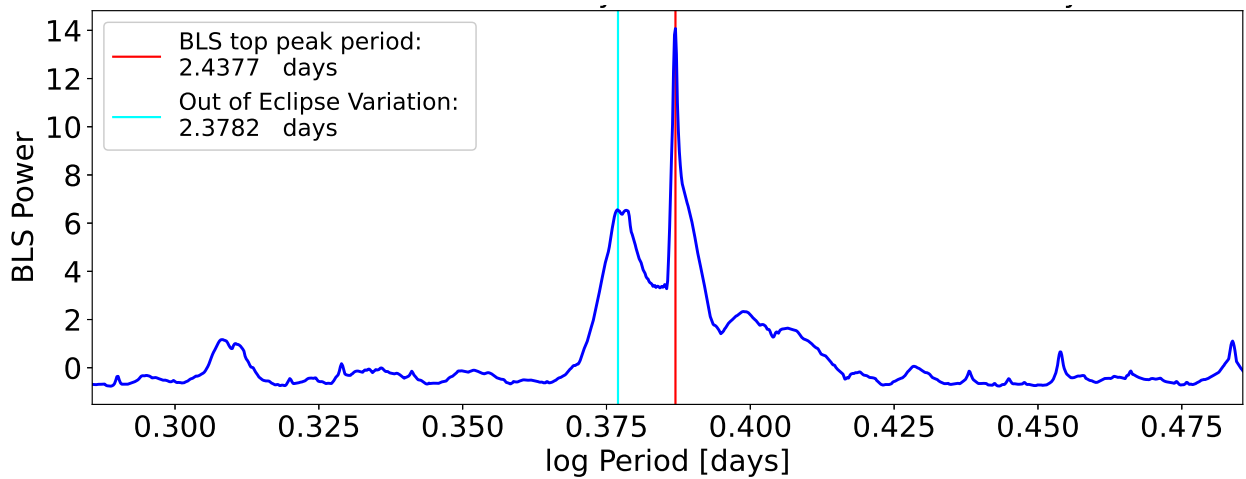}
    \caption{BLS periodogram, showing the most prominent periodic features, corresponding to the eclipse period but also showing a similar but different period of 2.3782~d which we attribute to the stellar rotation period (see the text).}
    \label{fig:ooe_bls}
\end{figure}

We then utilized the \textrm{Astropy} implementation of the Box-fitting Least Squares (BLS) algorithm to estimate the period of the evident out-of-transit variations. We used a period range from 1.5 to 11~d  with 10000 steps in frequency and a transit duration range of 1 to 24~hr. In Figure~\ref{fig:ooe_bls}, we focused on a region of the BLS periodogram centered on 2.43~d and are able to identify the orbital period of 2.4377~d along with another prominent periodic feature corresponding to a period of 2.3782~d, which we attribute to the rotational period of one or both of the stars in the EB. 

\begin{figure}[!ht]
    \centering
    \includegraphics[scale=0.6,trim=0 0 0 25,clip]{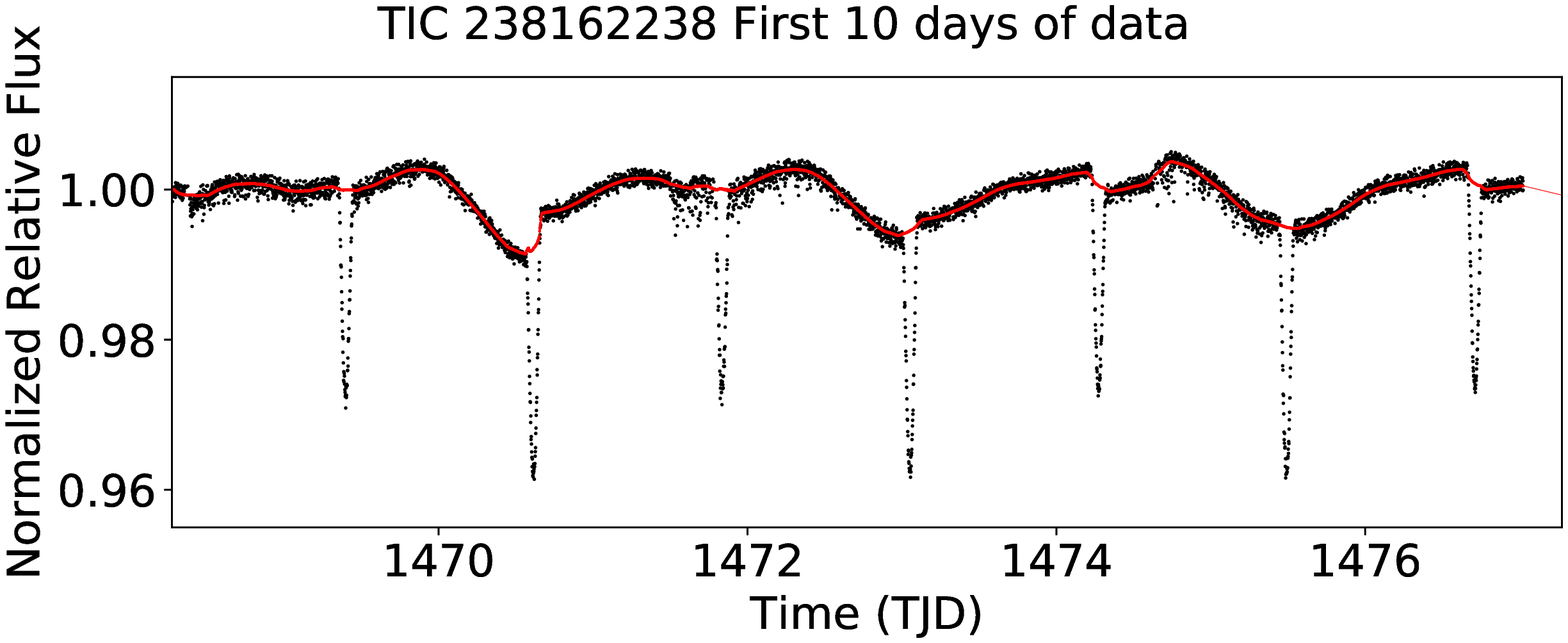}
    \caption{The first 10 days of the \textit{TESS} light curve spanning Sectors 6 to 8, shown in \textit{TESS} Julian Days (TJD). The black points are the \textit{TESS} data and the red line is the Biweighted Midcorrelation filter applied to the flux baseline. We used a window size of 6 hours which equals to 3 times the transit duration reported by BLS.}
    \label{fig:10day_tess_lc}
\end{figure}

\begin{figure}[!ht]
    \centering
    \includegraphics[width=0.9\linewidth,trim=0 0 0 25,clip]{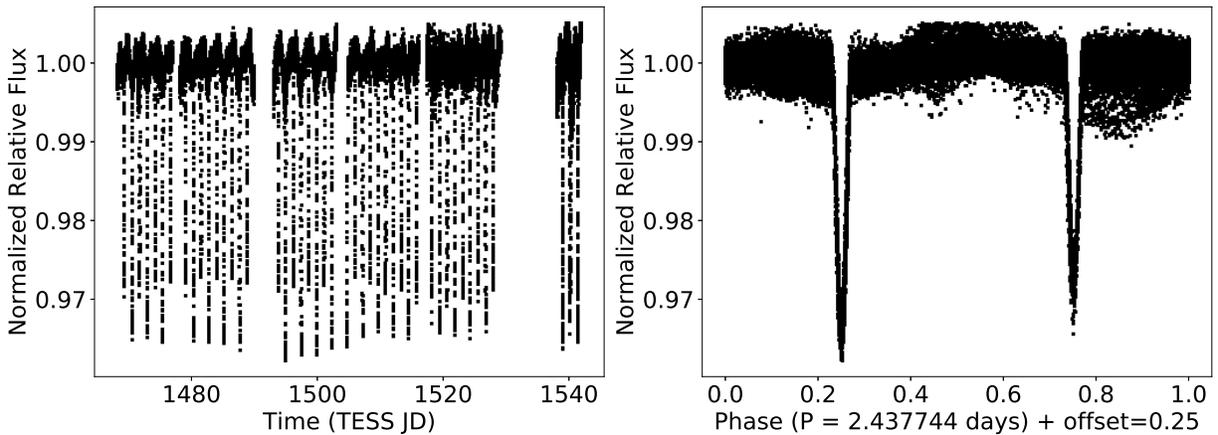}
    \caption{The phase folded \textit{TESS} light curve that was smoothed using a Biweighted Midcorrelation filter as described in \S\ref{sub:tess}. The out-of-transit variations seen in Figure \ref{fig:tess_lc} are removed in these data. }
    \label{fig:detrended_phasefold_tess_lc}
\end{figure}

To remove this out-of-eclipse variation, we then used a 24-hr window for our smoothing function. In Figure~\ref{fig:10day_tess_lc}, we show the first 10 days of our \textit{TESS} light curve with the trend line that is fit to the flux baseline in red. Since the Biweighted Midcorrelation filter is median-based, it is less sensitive to outliers and ignores the transit events. In Figure~\ref{fig:detrended_phasefold_tess_lc}, we show the final phase-folded \textit{TESS} light curve on the orbital period estimated above from the \textit{TESS} light curve. We report a final system ephemeris combining all of the available light curve data in Section~\ref{sec:ephem}. 

\subsection{Spectroscopic Observations}\label{sec:spectro_obs}

Spectroscopic Observations of the HD~54236 were acquired from the Wide Field Spectrograph (WiFeS) and echelle spectrograph on the ANU 2.3~m telescope from the Australian National University\footnote{http://rsaa.anu.edu.au/observatories/siding-spring-observatory}, and, the Goodman High Throughput Spectrograph \citep{Clemens:2004}, installed on the f/16.6 Nasmyth platform of the 4.1~m SOuthern Astrophysical Research (SOAR) telescope. In what follows, each instrument and the analysis of its spectroscopic observations are described.

\subsubsection{ANU 2.3m Spectroscopy}\label{sub:anu}
We obtained a single 100-sec exposure of HD~54236A using WiFeS \citep{Dopita:2007} on the ANU 2.3~m telescope at Siding Spring Observatory, Australia. WiFeS is an image slicer integral field spectrograph, where the B3000 grating and the RT560 dichroic were selected for our observations, giving the spectral coverage of 3500--6000\AA\ at a resolution of $\lambda / \Delta \lambda = 3000$. The object spectrum was extracted and reduced with the 
IRAF\footnote{IRAF is distributed by the National Optical Astronomy Observatory, which is operated by the Association of Universities for Research in Astronomy (AURA) under cooperative agreement with the National Science Foundation.} packages \textrm{CCDPROC} and \textrm{KPNOSLIT}, with the wavelength solution provided by a Ne-Ar arc lamp exposure taken on the same night. The spectrum is flux calibrated against exposures of spectrophotometric standard stars from \citet{Hamuy:1994}, employing techniques described in \citet{Bessell:1999}. 
The resulting flux-calibrated spectrum had signal-to-noise of S/N = 300 per resolution element. 

\begin{figure}[!ht]
\begin{center}
\includegraphics[scale=0.8,trim=0 0 0 191,clip]{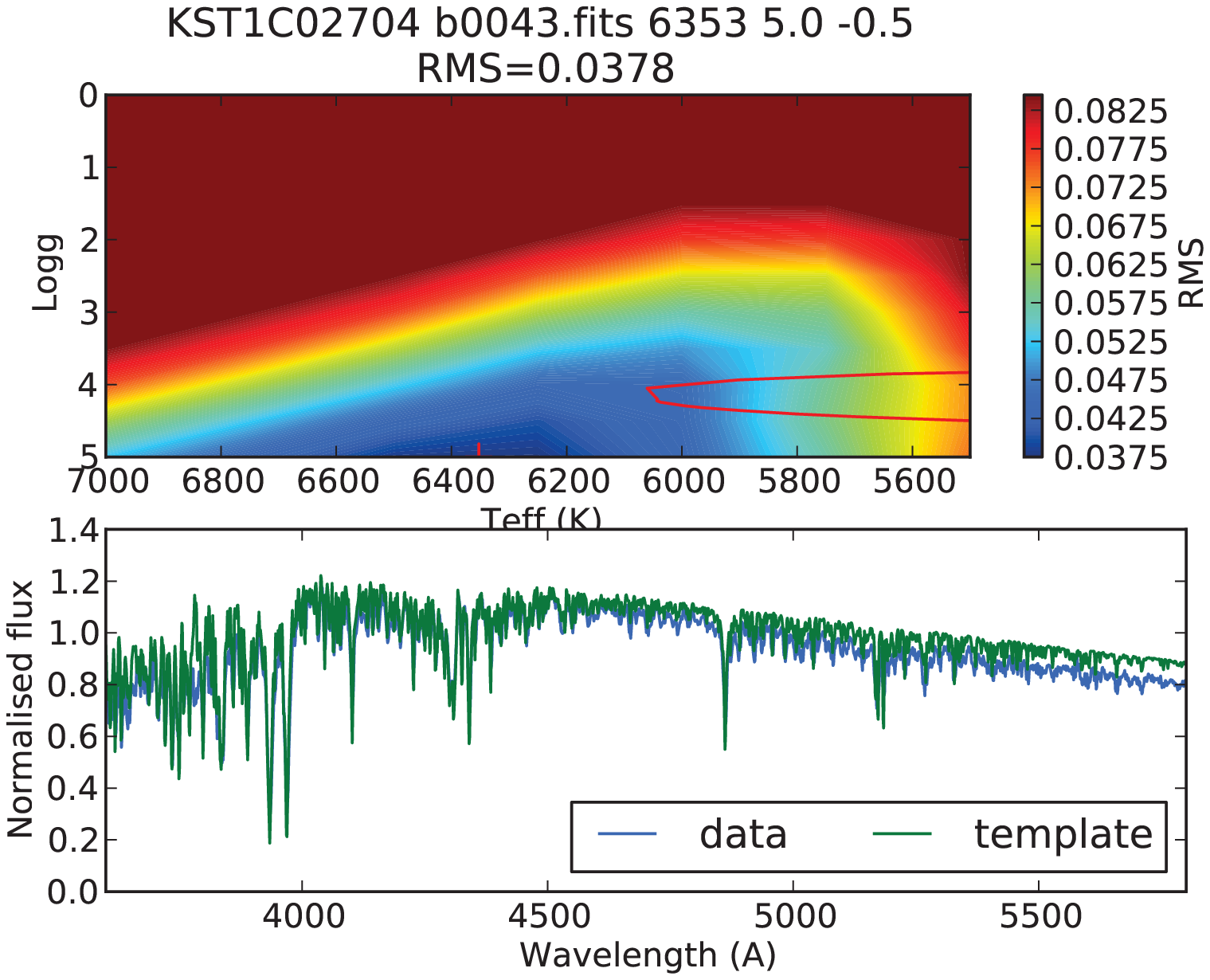} 
\includegraphics[scale=0.8,trim=0 0 0 191,clip]{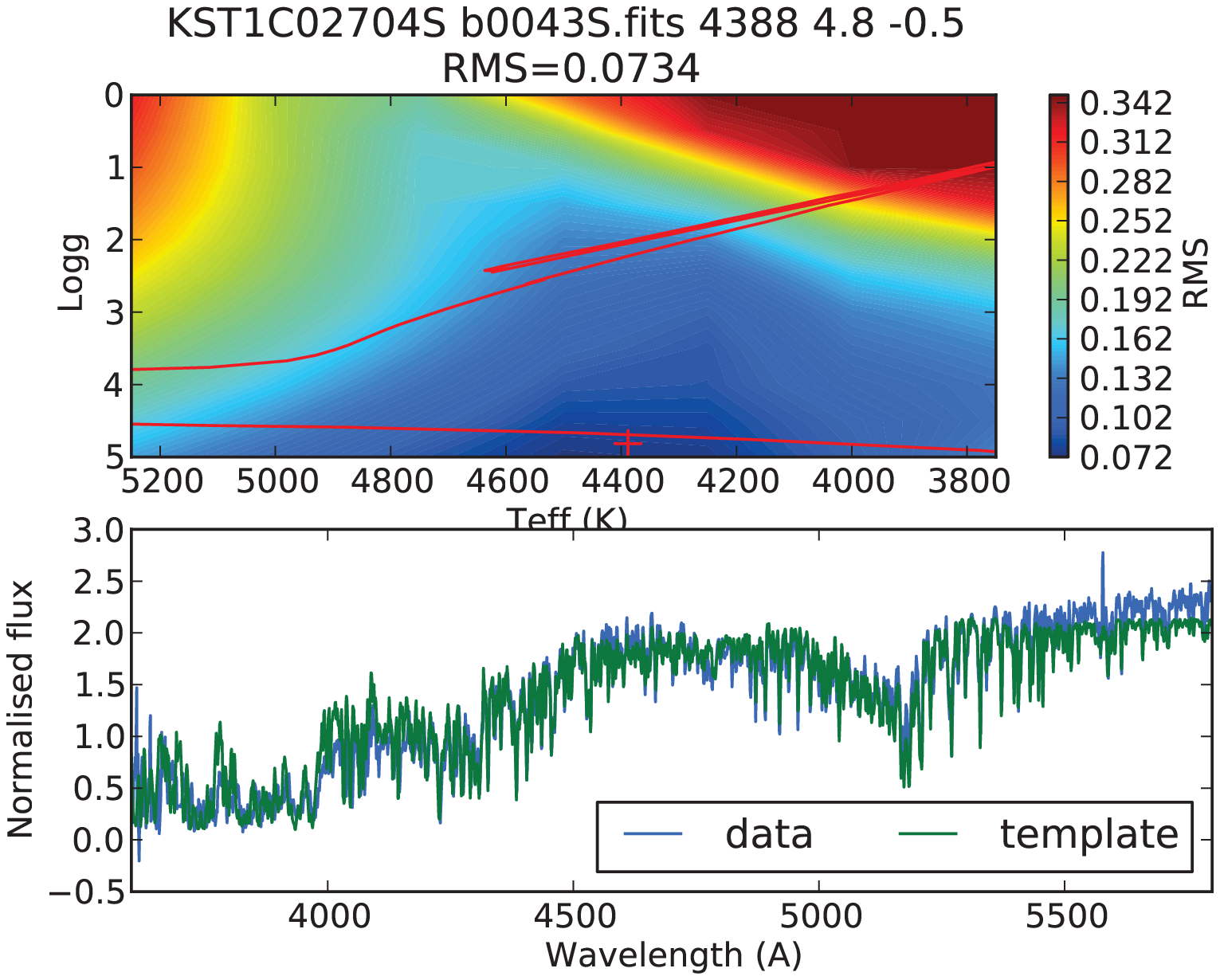}

\caption{\textit{Top:} Spectral analysis of the stellar parameters of HD~54236A, the eclipsing binary. \textit{Bottom:} Spectral analysis of the stellar parameters of the tertiary member, HD~54236B. The observed spectra for both HD~54236A and HD~54236B are shown in blue, while their respective templates modeled in green. Discrepancies between the target spectra and its template are largely due to differences in flux calibration, however, spectral matching is done based on spectral features rather than overall shape. 
\label{fig:Aparams}}
\end{center}
\end{figure}

The reduced object spectrum was fitted to a grid of synthetic spectra from \citet{Munari:2005}, spaced at intervals of 100~K in \teff, 0.5~dex in \logg, 0.5~dex in [Fe/H], and 0.02 magnitudes in $E(B-V)$. Discrepancies between the object spectrum and the synthetic template are largely due to differences in the flux calibration between them but prevent no analytical obstacle as spectral matching is done based on spectral features, rather than its overall shape.  We weighted regions sensitive to \logg\ variations preferentially, including the Balmer jump, the MgH feature at 4800\AA, and the Mg~b triplet at 5170\AA. Details of the data reduction and spectral fitting process can be found in \S~3.2.1 of \citet{Penev:2013}. 
The resulting derived atmospheric parameters for HD-54236A are 
\teff\ = $6350 \pm 100$~K, \logg\ $= 5.0 \pm 0.3$, and [Fe/H] $= -0.5 \pm 0.5$ (see Figure~\ref{fig:Aparams}, top panel).
Note that this analysis assumes that HD~54236A is a single star, when in fact it is a binary. We return to a discussion
of this in \S~\ref{sub:age}. 

We also observed the fainter CPM tertiary companion star HD~54236B. 
The analysis is displayed, as for HD~54236A, in Figure~\ref{fig:Aparams} bottom panel, and yields the following stellar parameters: 
\teff\ = $4400 \pm 200$K, \logg\ = $4.8 \pm 0.35$, [Fe/H] = $-0.5 \pm 0.44$ \citep{Bayliss:2013}.
%
%
In addition, from this spectrum we have verified the Li~EW of HD~54236A reported by SACY (120~m\AA). 
We obtain an EW of 140$\pm$20~m\AA\ for the Li line at 6708~\AA, representing the blended EW of both stars in the HD~54236 EB.

The WiFeS spectrograph was also used to measure the radial velocities of the eclipsing stars. One spectrum was obtained on HJD 2456104.838 using WiFeS at medium resolution $(\lambda / \Delta \lambda = 7000)$, where both components of the spectroscopic binary were resolved. The observation and data analysis process follows \citet{Penev:2013}. The velocities were measured via cross correlation as described in the next section, and are reported along with the full set of velocities described below in Table~\ref{tab:rv}. 

\subsubsection{High-resolution Spectroscopy: Radial velocities} 
Nine high-resolution observations of HD~54236A were obtained using the ANU 2.3~m echelle spectrograph, at a resolution of $\lambda / \Delta \lambda \approx 23000$, velocity dispersion of 4.0~km ${\rm s}^{-1} {\rm pixel}^{-1}$, in the spectral range 4200--6700\AA, over 20 echelle orders. The data was reduced with the IRAF package \textrm{CCDPROC}, extracted and normalized using \textrm{ONEDSPEC}. The wavelength solution was provided by Th-Ar arc lamp exposures that bracketed each science exposure. A standard 950~sec exposure of the target yields a signal-to-noise of S/N = 40 per resolution element. The instrument setup, observations, and data reduction process are detailed in \citet{Zhou:2014}. 

\begin{figure}[!ht]
\begin{center}
\includegraphics[scale=1.0]{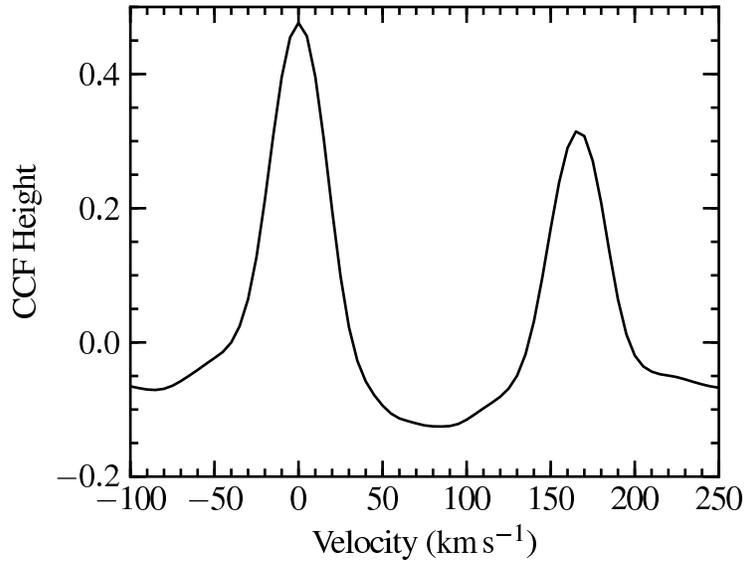} 
\caption{Cross-correlation function for the spectrum taken on HJD 2456057.93945226, with the ANU 2.3~m echelle at $R=23000$. The CCF shown is the average CCF of the cross correlations from all echelle orders, excluding those with very low cross-correlation heights. The cross-correlations are performed against RV standard star exposures taken on the same night. For clarity, the highest peak (presumed to correspond to higher mass, primary component in the eclipsing binary, has been shifted in this figure to be at 0~km$s^{-1}$.
\label{fig:ccf}}
\end{center}
\end{figure}

We cross-correlated the object spectra against a series of radial velocity standard star spectra taken during twilight each night. The cross-correlation peaks for the two spectroscopic components of the object were resolved in the observations, we were therefore able to extract radial velocities for both components in each observation by fitting Gaussians to the two cross-correlation peaks. We take the stronger peak to represent the (more massive) primary component. An example cross-correlation function (CCF) for the observation on HJD~2456057.93945 is shown in Figure~\ref{fig:ccf}.

The radial velocity is calculated as the mean over all orders not affected by telluric contamination, weighted by their respective signal-to-noise ratio; the scatter in the measurements over the orders gives the error in the radial velocity measurement for that epoch. 
The complete set of primary and secondary radial velocities determined from these observations and are summarized in Table~\ref{tab:rv}. 

\begin{table}[!ht]
\centering
\caption{RV Observations of HD~54236 A}\label{tab:rv}
\begin{tabular}{|c|c|c|r|}
\tableline
HJD & Primary RV & Secondary RV & Source \\
(UTC) & (km s$^{-1}$) & (km s$^{-1}$) & \\
\tableline
2456057.93945 &  $-88.7 \pm 1.3$ & $  76.9 \pm 0.9$ & ANU 2.3m \\
2456057.95085 &  $-90.6 \pm 1.4$ & $  78.9 \pm 1.1$ & ANU 2.3m \\
2456058.92243 &  $ 21.7 \pm 1.2$ & $ -46.3 \pm 1.1$ & ANU 2.3m \\
2456059.92704 &  $  6.6 \pm 1.5$ & $ -35.2 \pm 0.8$ & ANU 2.3m \\
2456061.96015 &  $ 84.9 \pm 1.4$ & $-109.1 \pm 1.1$ & ANU 2.3m \\
2456086.48771 & $  63.4 \pm 1.0$&  $-90.4 \pm 1.0$  & SOAR 4.1m \\
2456087.85052 &  $-55.9 \pm 2.0$ & $  36.9 \pm 1.9$ & ANU 2.3m \\
2456087.86683 &  $-54.3 \pm 1.6$ & $  30.1 \pm 2.7$ & ANU 2.3m \\
2456088.84221 &  $ 75.9 \pm 1.2$ & $-102.7 \pm 1.0$ & ANU 2.3m \\
2456088.85501 &  $ 73.4 \pm 1.2$ & $-101.7 \pm 1.1$ & ANU 2.3m \\
2456104.83853 &  $-68.8 \pm 1.1$ & $  61.3 \pm 1.6$ & ANU 2.3m \\
\tableline
\end{tabular}
\end{table}

\subsubsection{SOAR Spectroscopy}\label{sub:soar}
HD~54236 A \& B were observed on UT 20120605 and UT 20120607 using the Goodman High Throughput Spectrograph \citep{Clemens:2004}, installed on the f/16.6 Nasmyth platform of the 4.1m SOuthern Astrophysical Research (SOAR) telescope. The spectrograph used the 1200 l/mm grating in M5 mode, a 0.46-arcsecond wide slit, a GG-495 blue-blocking filter, imaged onto a 4096$\times$4096 Fairchild CCD detector, having square 15$\mu$m pixels (0.15 arcsecs/pixel). This setup yields a central wavelength of $\simeq$ 6890\r{A}, a wavelength range of $\simeq$ 6275--7495\r{A}, a spectral dispersion of 0.304\r{A} per pixel and a three-pixel resolving power of R$\simeq$7000 at 6400\r{A}. The FWHM of cross-correlated CuHeAr arc lines is 4.7279 pixels or 62.58~km~s$^{-1}$. Using this setup, HD~54236 A and B were separately observed for exposure times of 300- and 900-seconds, respectively, on UT 20120605, and 300- and 600-seconds, respectively, on UT 20120607, resulting in spectra having maximum S/N near to H$\alpha$ of $\simeq$ 350 for HD~54236A and 100 for HD~54236B.

For each target, we measured heliocentric radial velocities and \ion{Li}{1} 6708\r{A} equivalent widths. CCD instrumental effect removal (such as de-biasing and flat-fielding), as well as the extraction of wavelength calibrated spectra, was performed using standard IRAF procedures. Heliocentric radial velocities were determined relative to the International Astronomical Union standard stars HR~6468, HR~6859, HD~120223 and HD~126053. Cross-correlation of the velocity standards against one other shows that the zero-point relative to the standard system is accurate to $\sim 0.2$~km~s$^{-1}$. The CCF is the average from all echelle orders, excluding those with very low cross-correlation heights and are performed against RV standard star exposures taken on the same night. The radial velocity standard stars on these nights shows that the system is stable to $\simeq$0.5~km~s$^{-1}$ over the course of the night. 

The radial velocities for the eclipsing components of HD~54236A are reported with the other radial velocity measurements in Table~\ref{tab:rv}. For the single star HD~54236B, we obtain a systemic velocity of $-12.0 \pm 0.3$~km~s$^{-1}$. As shown below, this velocity is 
consistent with the systemic velocity of the HD~54236A eclipsing binary system. 

\begin{figure}[!ht]
\centering
\includegraphics[scale=0.5,angle=-90,trim=140 0 25 200,clip]{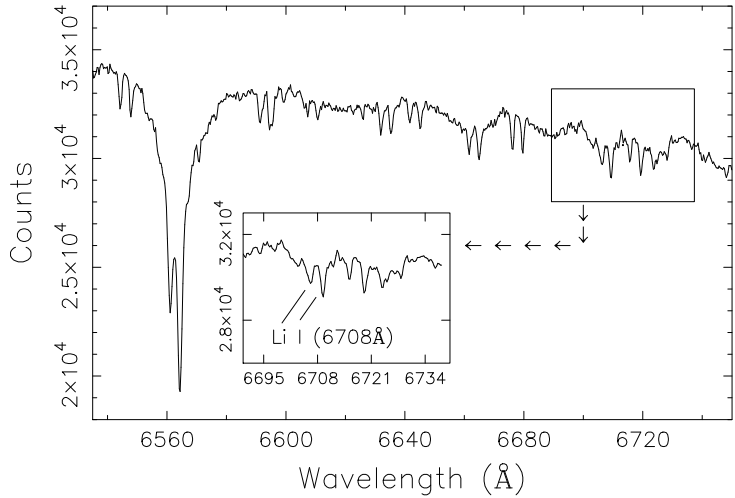} 
\caption{Region of the Li line from the SOAR spectrum.
\label{fig:soarli}}
\end{figure}

In order to measure the EW of the \ion{Li}{1} 6708~\r{A} resonance lines, the Goodman spectra were first trimmed to 6620--6750~\r{A} and normalized using a 7th-order spline function fit to the continuum. Equivalent widths were measured using both Gaussian fits and direct integration to check for consistency. The region of the spectrum around the Li line is shown in Figure~\ref{fig:soarli}. The measured Li EWs for the eclipsing stars in HD~54236A are $66 \pm 6$~m\AA\ for the primary star and $47 \pm 5$~m\AA\ for the secondary star. Note that these are the directly observed EWs; below we correct these EWs for the mutual dilution of the two stars' spectra when we transform these EWs into Li abundances. 

\section{Results}\label{sec:results}

With the above data in hand for the HD~54236 system, including photometric light curves and spectroscopic measurements of the component radial velocities and individual stellar spectra, in this section we determine the orbital parameters as well as constraints on the system age from consideration of the Li absorption and Ca~HK emission. To remind the reader, the system comprises three stars, two eclipsing components in HD~54236A and a single CPM star in HD~54236B. 

\subsection{System Ephemeris}\label{sec:ephem}
Analysis of all of the available light curve and radial-velocity data together (see \S\S\ref{sec:photo_obs}--\ref{sec:spectro_obs}), including in particular the KELT, PEST, and the exquisitely precise \textit{TESS} light curves, spanning a total time baseline of more than 10~yr, 
leads to the final orbital ephemeris determination of 
$P_{\rm orb} = 2.43742 \pm 0.00014$~d and HJD$_0$ = $2455199.5980 \pm 0.0021$. 

\subsection{Orbit Solution} \label{sub:orbit}
\begin{figure}[!ht]
\centering
\includegraphics[scale=0.4,angle=90]{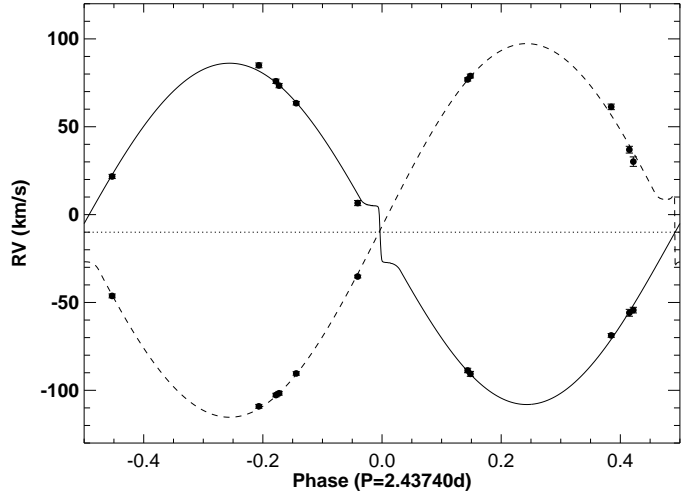}
\caption{RV solution for HD~54236A: The primary is modeled with the solid line, and the secondary with the dashed line.  
\label{fig:orbit}}
\end{figure}

We performed a simultaneous Keplerian orbit fit to the radial velocity measurements of the two eclipsing components of HD~54236A. We fit the ANU and SOAR radial velocities simultaneously, without attempting to introduce any possible systematic offset between the two. Thus, in total we fit 11 primary radial velocities and 11 secondary radial velocities which are listed in Table~\ref{tab:rv}. The resulting orbit solution is shown in Figure~\ref{fig:orbit} and summarized in Table~\ref{tab:orbit}. 
The orbital eccentricity is consistent with zero, which is not surprising given the very short orbital period, hence for the remainder of our analysis we assume a circular orbit ($e\equiv 0$). 

\begin{table}[!ht]
\centering
\caption{Orbital Solution of HD~54236 A}\label{tab:orbit}
\begin{tabular}{|c|c|c|c|}
\tableline
$q$ & $e$ & $v_\gamma$ & $a \sin i$ \\ 
\tableline
$0.9132 \pm 0.012$ & $0.009 \pm 0.008$ & $-10.0 \pm 0.5$ \kms & $9.810 \pm 0.080$ \rsun \\ 
\tableline
\end{tabular}
\end{table}

\subsection{\textrm{PHOEBE} Light Curve Model}

The PHysics of Eclipsing BinariEs, \citep{Phoebe:2011, Phoebe:2016} commonly referred to as PHOEBE is an EB modeling code which enhances upon the Wilson-Devinney \citep{Wilson:1971} method with improved model fidelity. Functions are available through \textrm{PHOEBE} that provide generation of highly characterized synthetic light curves given orbital and stellar parameters. 
We invoke PHOEBE to produce synthetic light curves for model fitting to our detrended TESS light curve, adopting the orbital parameters determined from the orbit solution (Section~\ref{sub:orbit}), such that the principal fitting parameters are the ratio of effective temperatures ($T_{\rm ratio}$), the sum of the stellar radii ($R_{\rm sum}$), and the orbital inclination ($i$). This is appropriate given the circular orbit, where to first order $T_{\rm ratio}$ is set by the relative eclipse depths and $R_{\rm sum}$ is set by the eclipse durations, moderated by $i$. 

We use a Genetic Algorithm (GA) to converge to a model light curve solution. Previous work \citep[see, e.g.,][]{Metcalfe:1999} has shown GA techniques to be robust for model fitting eclipsing binary light curves and thus inspired a modified version we utilize here. For computational efficiency we downsample the TESS 2-min cadence light curve by a factor of 15 and run \textrm{PHOEBE} using the Legacy backend, which is markedly faster.
The fitting algorithm starts with the generation of an initial random population of 500 model parameter sets [$T_{ratio},R_{\rm sum},i$] in our parameter space, each of which is encoded as a single object representation referred to as a gene. For each model we set the effective temperatures by apportioning the spectroscopically determined average ($6350 K$) according to the flux-weighted sum of the two stars and the current iteration of $T_{\rm ratio}$. 
For each gene a corresponding model light curve is computed in \textrm{PHOEBE}, and the $\chi^2$ between the observed and model fluxes is calculated. 

To thoroughly explore our parameter space we propagate this population over the course of 20 generations resulting in a total sample of 10,000 models. Iteration of genes from one generation (parent) to the next (child) is done using single point crossover. The members of each parent generation used in the crossover are chosen via roulette wheel until re-population is achieved. The probability of each selection is weighted by the reciprocal of the $\chi^2$ goodness-of-fit metric. To help maintain diversity between generations (i.e., to avoid local minima) we introduce a $10\%$ chance for random mutation within the constraints for the parameters at inception. 

\begin{figure}[!ht]
    \centering
    \includegraphics[width=0.7\linewidth]{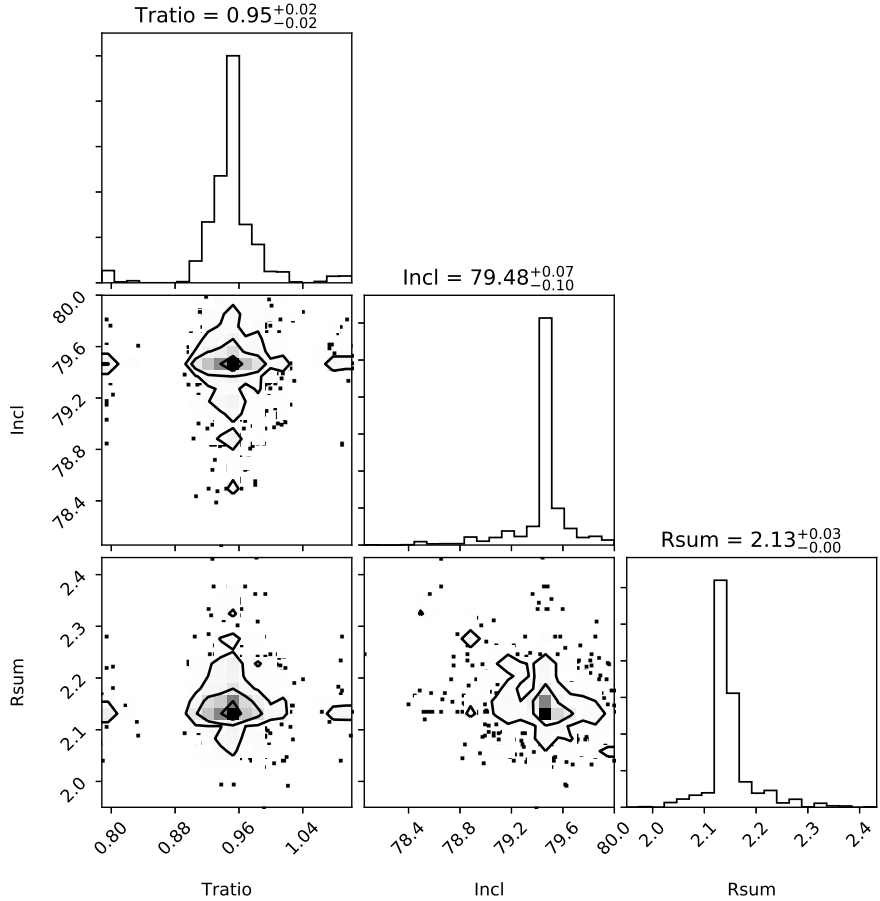}
    \caption{Corner plot of fitted parameters for cleaned sample of 2500 model light curves. Values atop distribution are the median values and the 0.16 \& 0.84 quantiles.} 
    \label{fig:corner}
\end{figure}

\begin{figure}[!ht]
    \centering
    \includegraphics[width=0.8\linewidth]{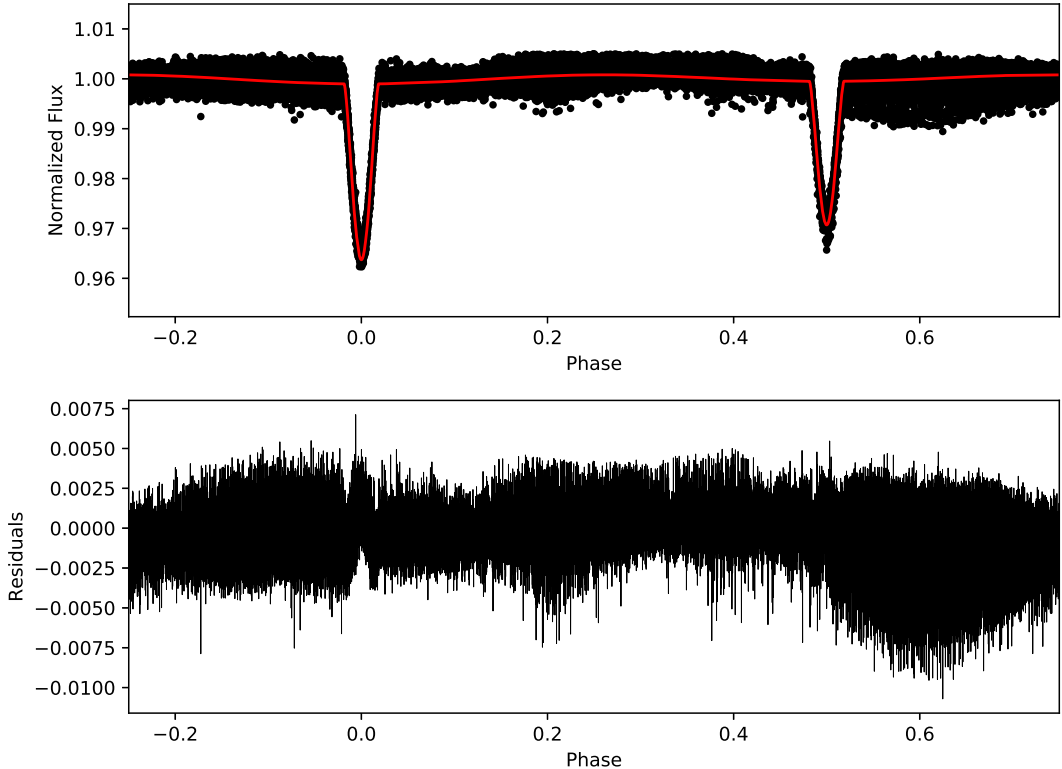}
    \caption{Light curve fitting for HD~54236A 
    using GA converged model solution.  
    There remain some small systematics in the residuals in eclipse, the amplitude of which are of order 1\% of the eclipse depths. The final stellar parameters from this solution are listed in Table~\ref{tab:fit}.}
    \label{fig:final_lcfit}
\end{figure}

After sampling is complete we discard the first 15 generations in order to prevent the lack of convergence in previous generations from contaminating our result. The fitted parameter distributions of the cleaned sample of 2500 models is illustrated in the corner plot (Fig.~\ref{fig:corner}). The distributions are gaussian-like in shape with well determined modal values. Thus, we determined the converged solutions as the mean values of these distributions and the uncertainties as the standard deviations. The final converged model is shown over our TESS light curve in Figure~\ref{fig:final_lcfit} as well as the residuals between the two. 
There is evidence of some minor systematics in the residuals of the primary eclipse with an amplitude of $\sim$0.003 relative flux units, however this is only $\lesssim$1\% of the eclipse depth of $\sim$0.035 relative flux units, which we deem acceptable. 

We use the fitted $T_{\rm ratio}$ together with the spectroscopically determined flux ratio of $3.0/4.5$, and solve Stefan-Boltzmann's Law to obtain $R_{\rm ratio}$, which with the fitted $R_{\rm sum}$ yields the final individual radii. 
Similarly, the fitted $T_{\rm ratio}$ together with the final $R_{\rm ratio}$ and the flux ratio yields the final individual temperatures.
Finally, using the final $i$ we obtain the stellar masses via the orbit-solution determined $a\sin i$ and $q$. The final stellar parameters 
are reported in Table~\ref{tab:fit}.

\begin{table}[!ht]
\centering
\caption{\textrm{PHOEBE} Fit Parameters and Solutions of HD~54236A}\label{tab:fit}
\begin{tabular}{|c|l|}
\tableline
$T_{\rm{eff,1}}$    & $6480\pm 103\ K$ \\ 
$T_{\rm{eff,2}}$    & $6155\pm 155\ K$ \\
$R_{1}$            & $1.128\pm 0.044\ R_\odot$ \\
$R_{2}$            & $1.021\pm 0.044\ R_\odot$ \\
$i$              & $79.4\pm 0.2^{\circ}$ \\
$M_{1}$            &    $1.179\pm0.029$~\msun \\
$M_{2}$            &    $1.074\pm0.027$~\msun \\ 
\tableline
\end{tabular}
\end{table}

\subsection{SED Analysis}

\begin{figure}[!ht]
    \centering
    \includegraphics[scale=0.475]{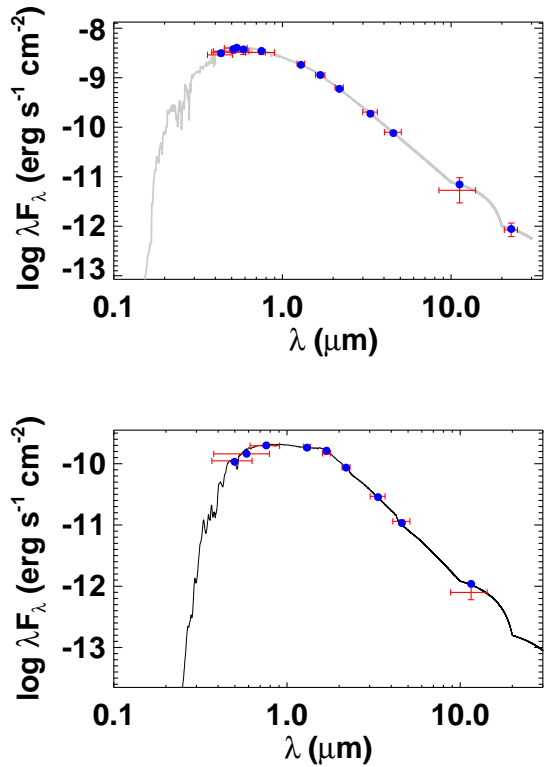}
    \caption{\textit{Top:} SED fit of HD~54236A \textit{Bottom:} HD~54236B SED fit. Red symbols are observed broadband fluxes, curves are the best-fit model atmospheres, and blue symbols are the model fluxes for comparison to the observed fluxes.}
    \label{fig:sed}
\end{figure}

The radii and effective temperature solutions from the \textrm{PHOEBE} analysis (Table~\ref{tab:orbit} were used in Spectral Energy Distribution (SED) modeling, fitting only for extinction and distance. This is done to ensure that our \teff\ solutions correctly reproduce the observed shape of the SED, and that the inferred distance from this analysis corresponds to the known \textit{Gaia} distance. From Figure~\ref{fig:sed} the SED fits for HD~54236A (top) and HD~54236B (bottom) show a very good agreement with our \teff\ determinations. The \textit{Gaia} distance to HD~54236 is 131.8 $\pm$ 0.5~pc, the best-fit distance from the SED analysis is in agreement at 136 $\pm$ 6~pc, providing a strong validation of the temperatures and radii that we have determined for both stars in the EB.

For the companion star, HD~54236B, the \textit{Gaia} distance was adopted in fitting the SED to solve for the stellar radius. The stellar radius combined with the $\log g$ yields an estimate for the mass. For the radius we find $0.65 \pm 0.06$~\rsun\ and a mass estimate of $0.65 \pm 0.11$~\msun. These are fully consistent with the empirical relations of \cite{Torres:2012} who give a similar but more precise mass estimate of $0.67 \pm 0.04$~\msun. 

\subsection{System Age\label{sub:age}}

\begin{figure}[!ht]
\centering
\includegraphics[scale=0.5]{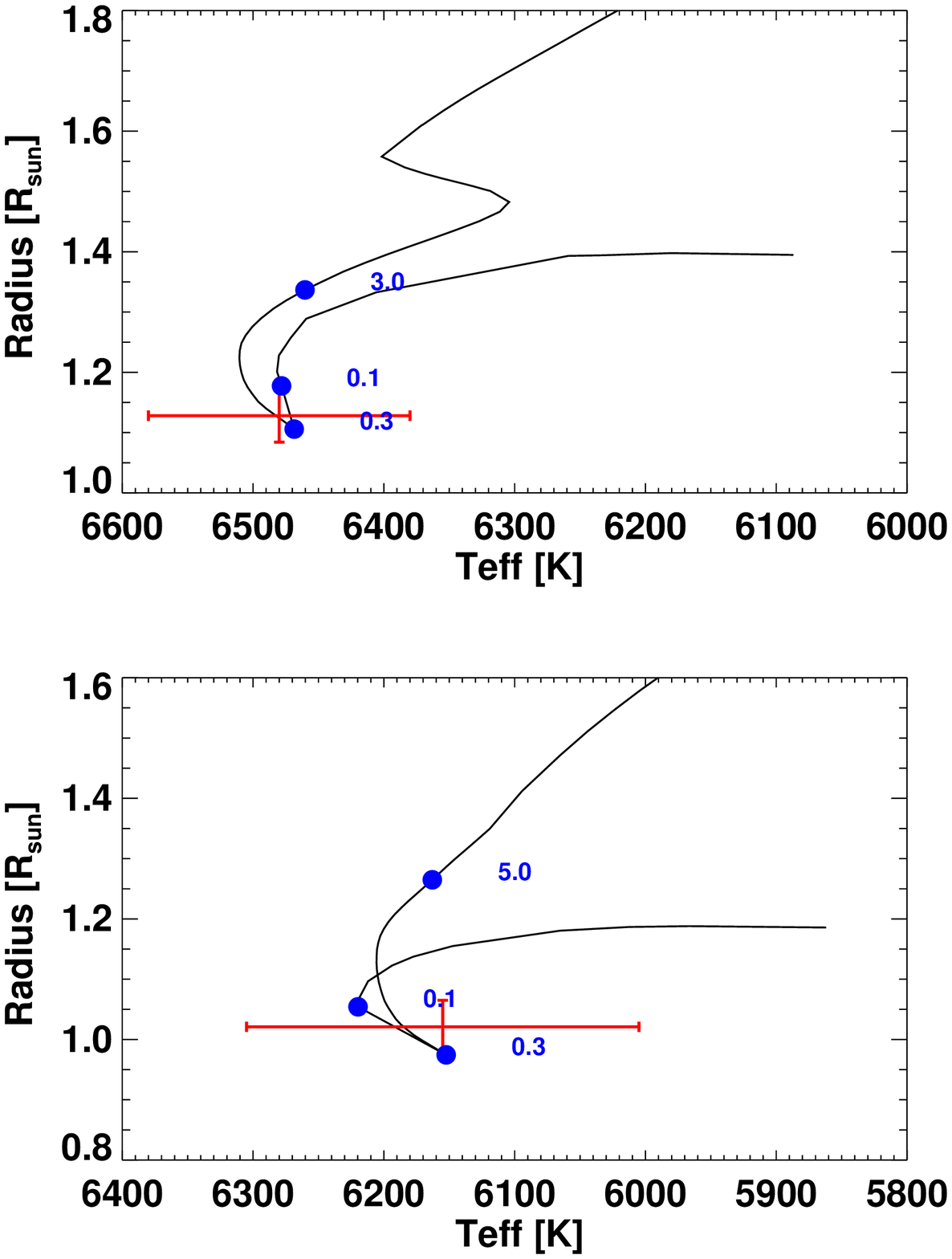}
\includegraphics[scale=0.5,angle=90]{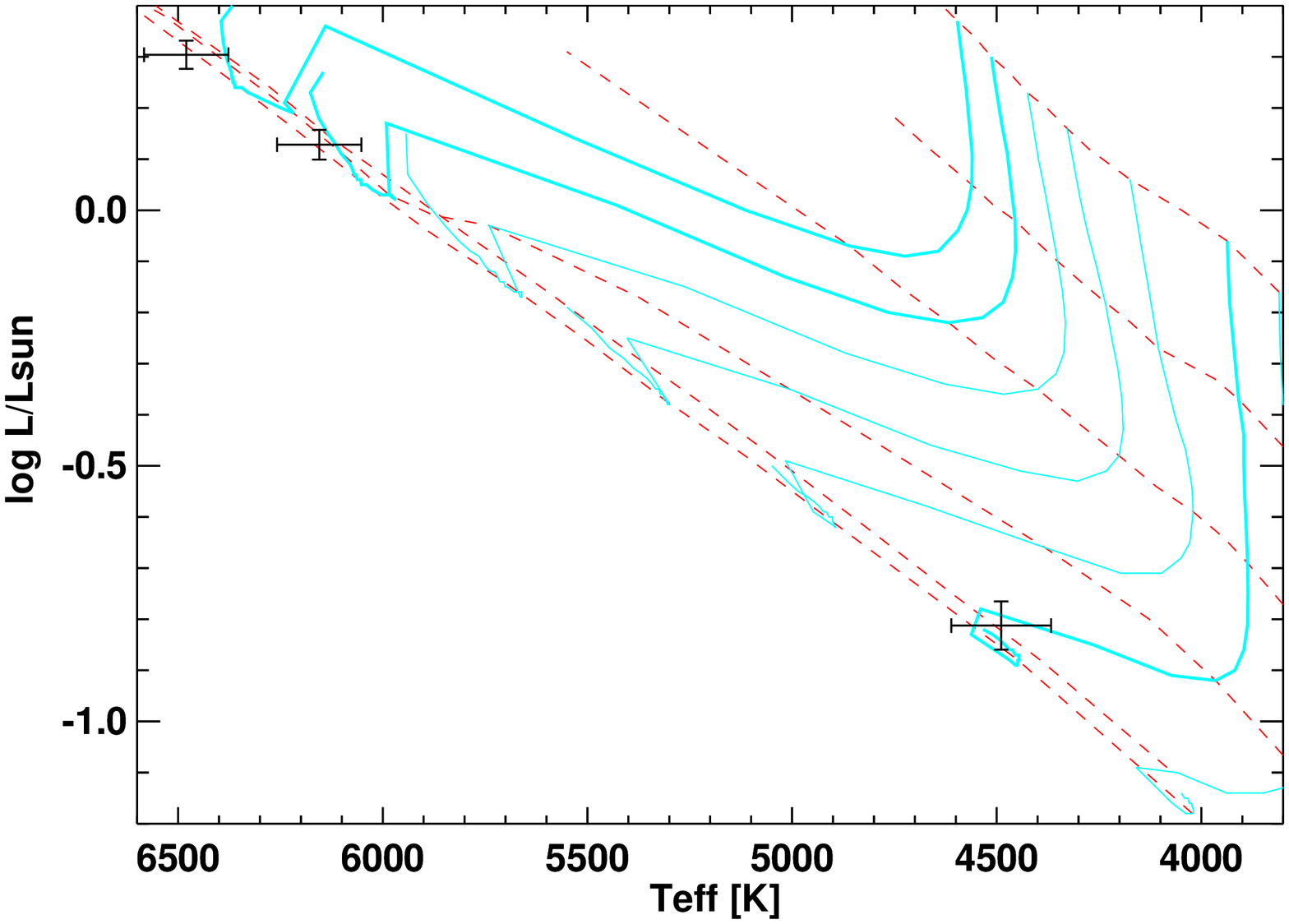}
\caption{\textit{Top:}\teff-Radius diagram for the primary star of the HD~54236A system. \textit{Middle:} \teff-Radius diagram for the secondary star of the HD~54236A system. Both are in the \teff\ vs Radius plane and  show the Yonsei-Yale stellar evolutionary tracks \citep{Yi:2001}, which suggest an age for HD~54236A near the ZAMS (which occurs at an age of 300~Myr in these models) and roughly consistent with an age of 225$\pm$50~Myr as determined from other age diagnostics.  (Bottom:) H-R diagram showing all three components of the HD~54236 system (black symbols) compared to the PMS evolutionary models of \citet{Baraffe:1998}; shown are isochrones at ages of 1, 3, 10, 30, 100, and 300 Myr (red dashed lines) and tracks for masses of 0.6--1.2~\msun\ in increments of 0.1~\msun\ (blue curves).
\label{fig:hrd}}
\end{figure}

In order to establish an age estimate for the HD~54236 system, we consider two commonly used stellar chronometers, namely the abundance of Li and the strength of chromospheric activity in the \ion{Ca}{2}~H\&K lines. While the \ion{Ca}{2} H\&K activity indicator will be enhanced in the eclipsing pair HD~54236A because of their short orbital period and therefore likely enhanced rotation, we utilize the existence of the wide tertiary companion star, HD~54236B, to obtain a clean estimate of the \ion{Ca}{2}~H\&K activity age for the system. This analysis is carried out through the comparison of \ion{Ca}{2}~H\&K spectral features against those of similar effective temperatures from three different open clusters spanning the range of age estimates covered in \ref{sub:lithium}.   
Finally, in an attempt to establish association of the system with other nearby stars of similar age, we  consider the space motion of the HD~54236 system. For context, in Fig.~\ref{fig:hrd} we show the components of the HD~54236A system in the \teff--Radius plane compared to the Yonsei-Yale stellar evolutionary tracks \citep{Yi:2001} for the spectroscopically determined [Fe/H] of $-0.5$. We also represent all three stars---the eclipsing components of HD~54236A and the tertiary HD~54236B---in the H-R diagram plane compared to the PMS evolutionary models of \citet{Baraffe:1998}. All of which suggest an age near the ZAMS or slightly younger, consistent with the other age diagnostics discussed above. 

\subsubsection{Lithium}\label{sub:lithium}
The observed Li EWs for the three stellar components in the HD~54236 system are indicative of a relatively young system age intermediate between the youngest pre-main-sequence clusters and older main sequence clusters. For example, from the analysis of Li EW in young stars of various ages, \citet{Aarnio:2008} finds that young stars with K spectral type and ages of $\sim$50~Myr have Li EW of 100--200 m\AA, somewhat larger than what we observe in the HD~54236B star (spectral type late K). Therefore it is likely that the HD~54236 system has an age somewhat older than 50~Myr. To determine the Li inferred age of the system more precisely, we analyzed the observed Li EWs for the three stars in HD~54236 in the context of other young clusters. 

First, we corrected the observed EWs for the two eclipsing stars to account for the mutual dilution of the two stars' spectra in our combined light observations. 
Based on the stars' final \teff\ values from our \textrm{PHOEBE} modeling and SED fits, we converted the Li EW into Li abundances, $A(Li)$, using standard curve-of-growth techniques \citep{Soderblom:1993} and applying a NLTE correction \citep{Carlsson:1994}. 

\begin{figure}[!ht]
\begin{center}
\includegraphics[scale=0.8]{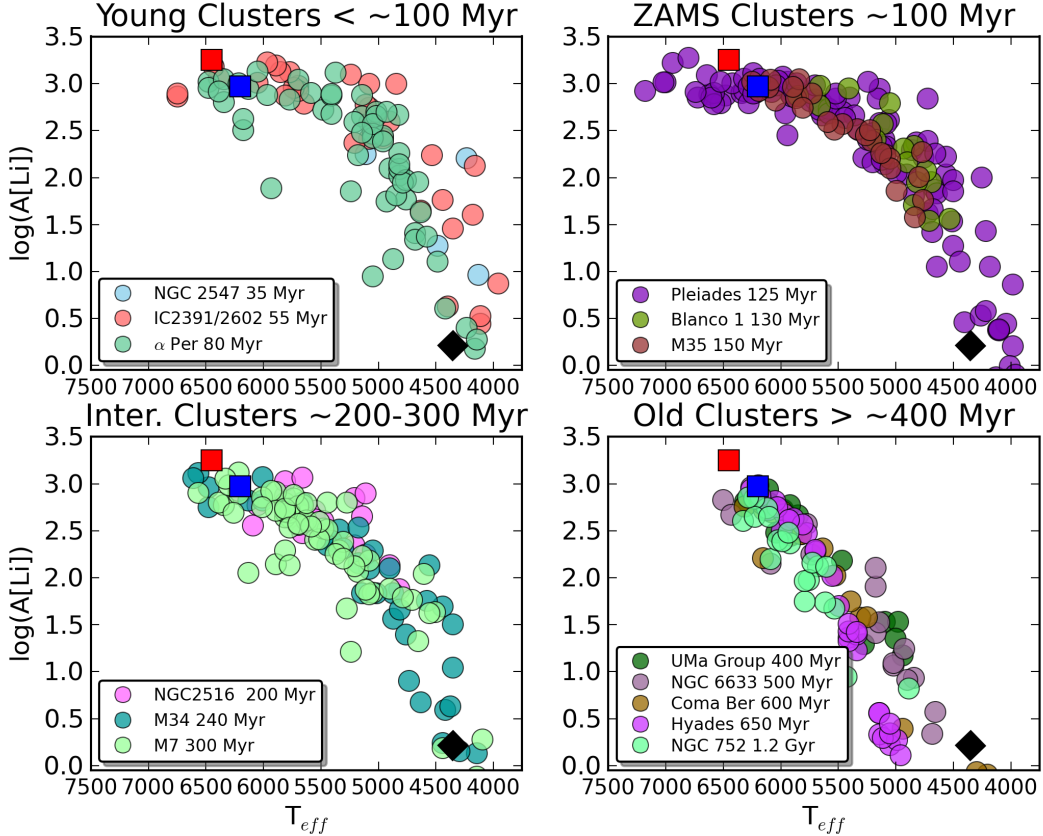} 
\caption{Lithium abundance analysis for a range of ages of clusters: Comparison of EW of lithium abundances from the HD~54236 system to several different clusters ranging in age from 35~Myr to 1.2~Gyr \citep[sample clusters and their parameters are from][]{Sestito:2005}.} From these comparisons we can see that the stellar components in the HD~54236 system (shown with red and blue squares for the primary and secondary stars in the eclipsing binary primary HD~54236A) are indicative of a relatively young age: between PMS clusters and older main sequence clusters. 
\label{fig:HD54236_Li}
\end{center}
\end{figure}

The sample from \cite{Sestito:2005} contains 22 clusters spanning a range of ages from $\sim$5-Myr to $\sim$6--8~Myr, for our comparison, we gathered $A(Li)$ for several of these open clusters ranging in age from 35~Myr to 1.2~Gyr and follow the same curve-of-growth and NLTE correction in our abundance calculation method for the HD~54236 system. The stellar censuses for our comparison clusters shown in \ref{fig:HD54236_Li} are as follows, in the young clusters with ages less than $\sim$ 100~Myr: NGC~2547 contains 7 stars, IC~2391 has 6, IC~2602 contains 31 and $\alpha$~Per contains 39 stars. In the ZAMS clusters, with ages $\sim$ 100~Myr are the Pleiades with its seven sisters, Blanco~1 which consists of 17 stars, and M35 with a total of 27. The intermediate comparison clusters have ages of $\sim$200--300~Myr and are represented by NGC~2516, M34 and M7 or NGC~6457 which have 22, 38, and 24 stars respectively. The oldest clusters used for comparison were of ages greater than $\sim$400~Myr, these included the UMa Group with 14 stellar members, NGC~6633 with 22 stars, Coma~Ber with roughly 40 stars forming its distinctive {\tt V} shape, Hyades containing 14 stars, and NGC~752, the oldest comparison, having 5 members. The data set used in \citet{Sestito:2005} (and used for comparison in this work) are a collection of independent results from a variety of authors using a degree of models to investigate the evolution of Lithium abundances and their associated timescales.

From visual inspection of Fig.~\ref{fig:HD54236_Li}, it is clear that the three stars' $A(Li)$ generally follow the trend with \teff\ observed in young clusters with ages of 100--400~Myr. To constrain the likely age more precisely, we calculated the $\chi^2$ probability of the three stars' $A(Li)$ agreement with each of the comparison clusters shown in the figure. We obtain the following probabilities for the HD~54236 system Li abundances: 
P(age $<$ 100~Myr) = 0.03, 
P(age $\sim$100~Myr)= 0.09, 
P(age 200--300~Myr)= 0.40, 
P(age $>$ 300~Myr) $<$ 0.01. 
Therefore, statistically the pattern of $A(Li)$ with \teff\ observed for the three stars in HD~54236 is most consistent with an age of 200--300~Myr. However, a younger age of $\sim$100~Myr cannot be entirely ruled out on the basis of the Li abundances alone. 

\subsubsection{Ca~H~\&~K}\label{sub:CaHK}
The strength of the emission cores in the H~\&~K lines of \ion{Ca}{2} near 3950\AA\ are a commonly used tracer of chromospheric activity and thus, through the empirical rotation-activity relationship of FGK dwarfs, can be used as an age indicator. Given the short orbital period of the eclipsing pair in HD~54236A, it is a good assumption that the two stars have tidally interacted and therefore the stars are likely to have their current rotation governed not by the rotation-activity-age relation, but rather by tidal synchronization with the orbit. Indeed, we observe a roughly sinusoidal variation in the eclipsing system light curve at very nearly the orbital period (see \S~\ref{sec:photo_obs}), confirming that the stars likely rotate at very nearly the orbital period due to their close separation. Current rotation-activity-age relationships (in this case the age-$R'_{HK}$ relationship) are based on single-star angular momentum evolution, so this rules out using the HD~54236A system to determine the system age using chromospheric activity measures. 

Fortunately, the single K-dwarf star in the system (HD~54236B) \textit{can} be used for this purpose. 
We utilize the high S/N spectrum obtained with the ANU WiFeS spectrograph. Both because our spectrum was not absolutely flux calibrated and because the late spectral type of HD~54236B places it beyond the calibration of existing age-activity relationships \citep[e.g.][]{Wright:2011, Mamajek:2008},we have instead simply compared our observed spectrum with suitable late-type standards from several young clusters.

\begin{figure}[!ht]
\begin{center}
\includegraphics[scale=0.7,trim=0 10 0 30,clip]{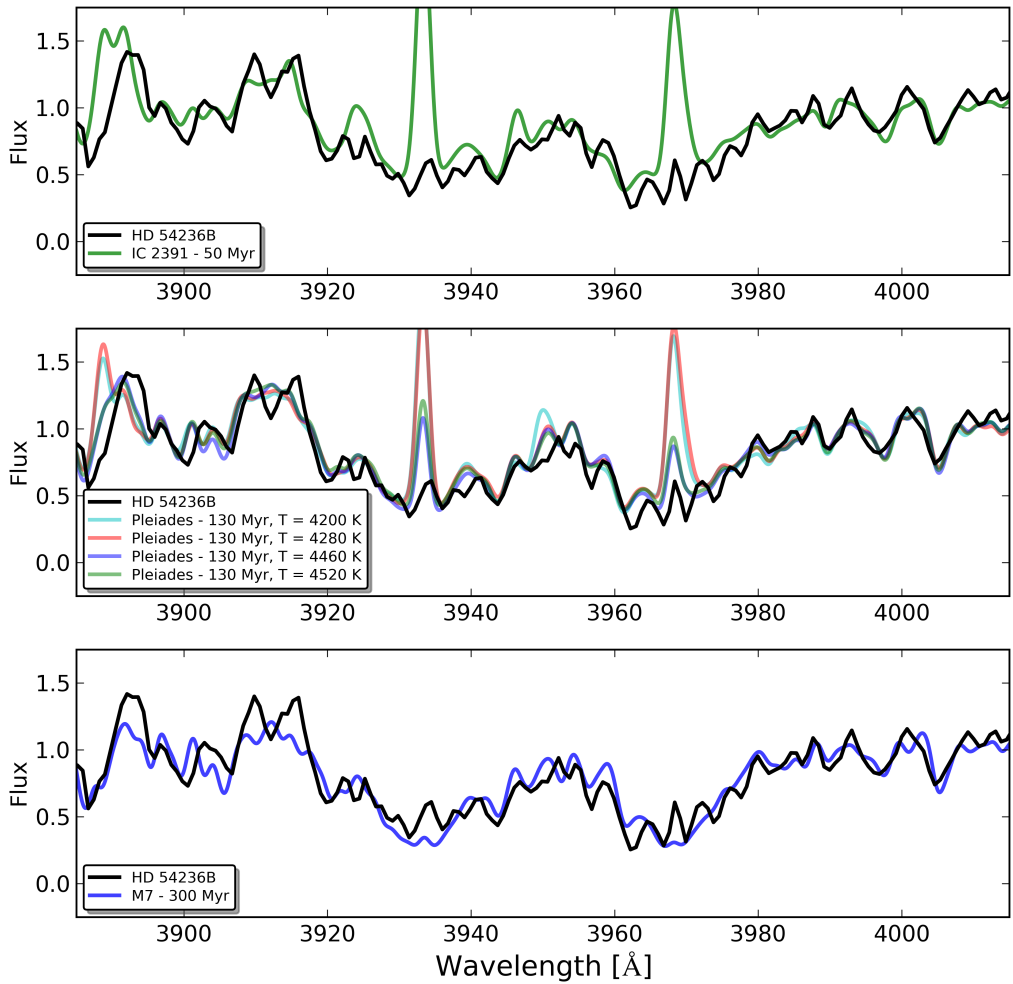} 
\caption{\ion{Ca}{2} H\&K analysis of the WiFeS spectrum for HD~54236B against stars with similar \teff\ from three different open clusters spanning the range of ages estimated from the Li abundances in \S \ref{sub:lithium}. Against the \ion{Ca}{2} H\&K analysis for HD~54236B (black line). \textit{Top:} comparison with IC 2391 (age $\sim$ 50~Myr). \textit{Middle:} comparison with Pleiades (age $\sim$ 130~Myr with a range of \teff\ from 4200~K - 5420~K). \textit{Bottom:} Comparison with M7 (age $\sim$ 300~Myr). The open cluster spectra were taken from Keck HIRES and VLT UVES public archives and convolved to match the resolution of the WiFeS observation of HD~54236B. Taken together, these comparisons strongly imply an age of HD~54236B older than the Pleiades ($\sim$130~Myr) and clearly younger than M7 ($\sim$300~Myr).
\label{fig:HD54236_CaII}}
\end{center}
\end{figure}

In Figure~\ref{fig:HD54236_CaII}, we plot the WiFeS spectrum of the \ion{Ca}{2} H\&K spectral region for HD~54236B against stars with similar effective temperatures ($4200K \leq T_{eff} \leq 4520K$) from three different open clusters spanning the same range of ages found in our analysis of the Li abundances in the previous section \ref{sub:lithium}.

The open cluster spectra were taken from the Keck HIRES and VLT UVES public archives and convolved down to match the lower resolution of the WiFeS observation. We performed a manual examination of the available spectra to confirm that the open cluster stars plotted are typical in their level of activity for their respective ages.

The agreement of the comparison open cluster spectra to the HD~54236B spectrum outside of the \ion{Ca}{2}~H\&K lines is good evidence that the selected open cluster stars are of appropriately similar spectral type. This is important because the change in level of activity in this age range is a fairly sensitive function of spectral type. For instance, at Pleiades age the \ion{Ca}{2}~H\&K emission cores transition from very strongly present for \teff $\lesssim$4300~K to modest (at this spectral resolution) for \teff $\gtrsim$4475~K, a change of only 175~K.  
Nevertheless, all of the Pleiades show significant \ion{Ca}{2} emission (even at the WiFeS resolution) at levels significantly higher than observed in HD~54236B. In contrast, there is also agreement between the observed mild emission in HD~54236B and the M7 comparison star (Bottom panel, \ref{fig:HD54236_CaII}). Together, these comparisons strongly imply an age for HD~54236B (and therefore of the HD~54236 system) that is older than the Pleiades ($\sim$130~Myr) and younger than M7 ($\sim$300~Myr). 

\subsubsection{Isochrone fitting}

Ancillary age estimates for HD~54236~B are fitted using photometry and MIST stellar evolutionary models by \texttt{MINESweeper} \citep[for extensive details on the fitting procedure, including priors used in the inference, see][]{Cargile:2019}. Using a Bayesian approach, \texttt{MINESweeper} estimates stellar parameters from MIST stellar evolutionary models \citep{Choi:2016}. \texttt{MINESweeper} employs multi-nested ellipsoid sampling \citep[due to its efficiency in sampling multi-modal likelihood surfaces like stellar isochrones;][]{Feroz:2019} and utilizes optimized interpolation following \citet{Dotter:2016}. Fitting \textit{Gaia} DR2, 2MASS, and UNWISE photometry for HD~54236B, \texttt{MINESweeper} and the latest MIST stellar evolutionary models yield an age estimate of $149^{+173}_{-77}$-Myr for HD54236~B, in agreement with our other age estimators outlined above. Being that HD~54236B is a single star, binarity is not an added complication to fitting and, having lower mass than its EB companion is also less evolved than the HD~54236A EB, therefore, HD~54236B's position on in HR diagram space is more sensitive to the system's age.

\subsubsection{Summary of Age Estimates}\label{sub:ages}
Our observation of the \ion{Ca}{2} H\&K mission in the HD~54236B K-dwarf star implies an age for HD~54236B (and therefore of the entire HD~54236 system) that is clearly older than the Pleiades ($\sim$130~Myr) and clearly younger than M7 ($\sim$300~Myr). This result is in excellent accord with that inferred from the Li abundances of all three stars in the HD~542436 system, for which we obtained a most likely age of 200--300~Myr. As the Li results more clearly rule out an age as young as 130~Myr and the \ion{Ca}{2} H\&K results more clearly rule out an age as old as 300~Myr, we adopt a most likely system age of $225 \pm 50$~Myr on the basis jointly of the Li abundances of all three stars and the \ion{Ca}{2} H\&K emission of the HD~54236B component. This adopted most likely system age of $225 \pm 50$~Myr is also in agreement with the latest MIST models projected age estimate of $225 \pm 50$~Myr. 

\section{Discussion: Membership, Stellar Strings, and AB~Dor}\label{sec:discussion}

\textit{Gaia} DR2 parallax estimates \citep{Bailer-Jones:2018} of HD~54236A puts it at a distance of 133~pc. The galactic space velocity is calculated using the \textit{Gaia} distance, proper motion and radial velocity estimates following \cite{Johnson:1987}. The $UVW$ velocities for the HD~54236 system are calculated using \textrm{astrolibR} \cite{Chakraborty:1995} and are listed in Table \ref{tab:UVW}. 

\begin{table}[ht]
\centering
\caption{Galactic Motions of HD~54236A}\label{tab:UVW}
\begin{tabular}{|c|c|c|c|}
\tableline
Parallax & $U$ & $V$ & $W$ \\
\tableline
$7.56\pm0.024$ & $9.07\pm0.61$~km$s^{-1}$ & $9.54 \pm 0.36$~km$s^{-1}$ & $-4.97\pm0.23$CPM~km$s^{-1}$ \\
\tableline
\end{tabular}
\end{table}

No membership to any population has been established HD~54236 to date. Indeed, the \textrm{BANYAN $\Sigma$} Multivariate Bayesian Algorithm \citep{Gagne:2018} reports 0\% probability of this source being associated with any of the 27 nearby moving groups that the algorithm considers, based on its \textit{Gaia} DR2 distance and proper motion estimates \citep{GaiaCollaboration:2018,Bailer-Jones:2018}.

Nonetheless as the number of known co-moving groups has grown significantly in the recent years, we search for other possible candidate populations from which HD~54236A may have originated. In particular, we perform a cross-match against \citet{Kounkel:2019,Kounkel:2020} catalog, searching for the nearest moving group of which this source may be the most likely member. Examining the sources located closer than $\pi>6$ mas, with position within 10 degrees, and having proper motions 8 mas yr$^{-1}$ of the source, only one group appears to be close, Theia~301, making it the most likely progenitor. 

\begin{figure}[!ht]
\epsscale{1.1}
 \centering
\plottwo{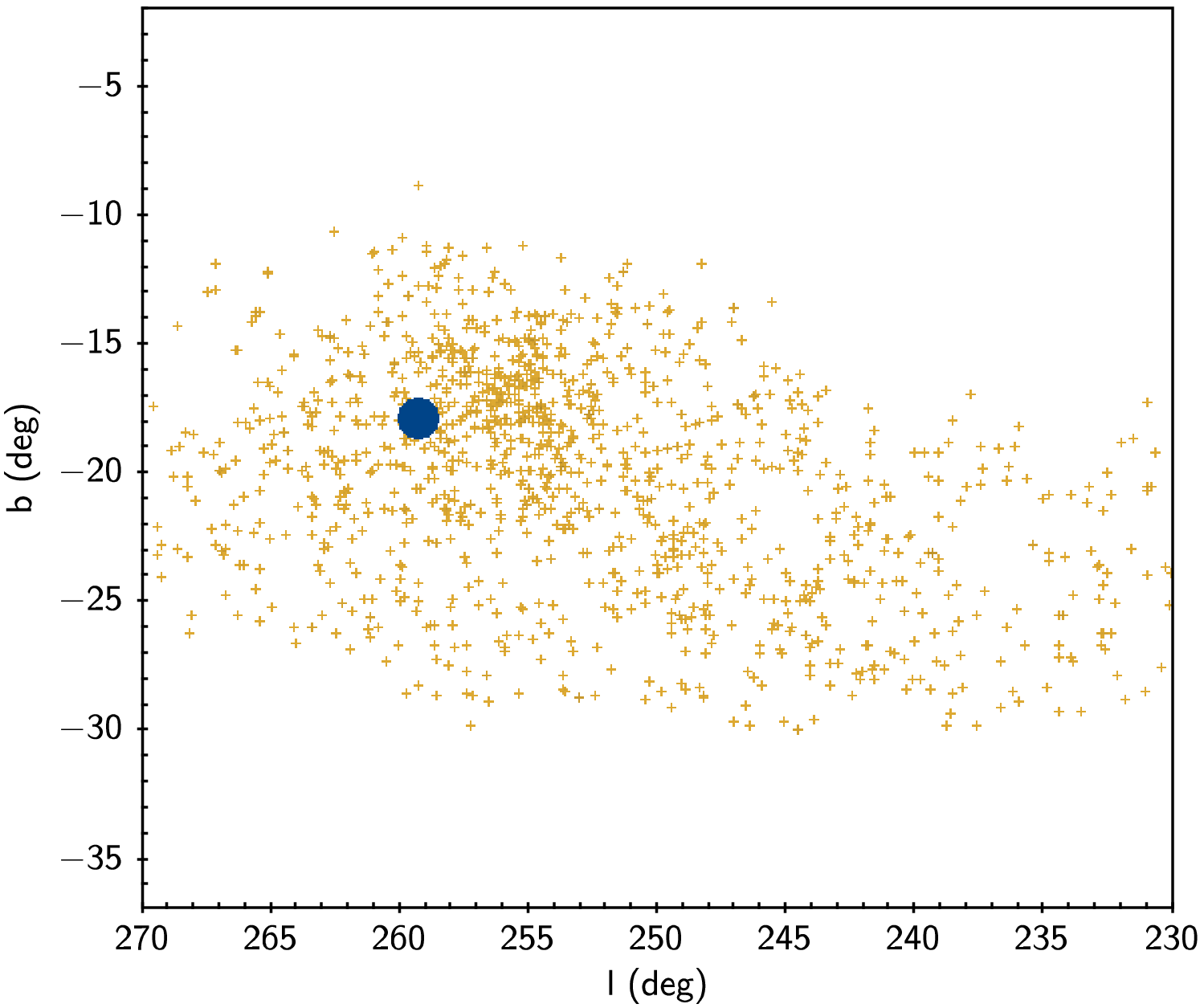}{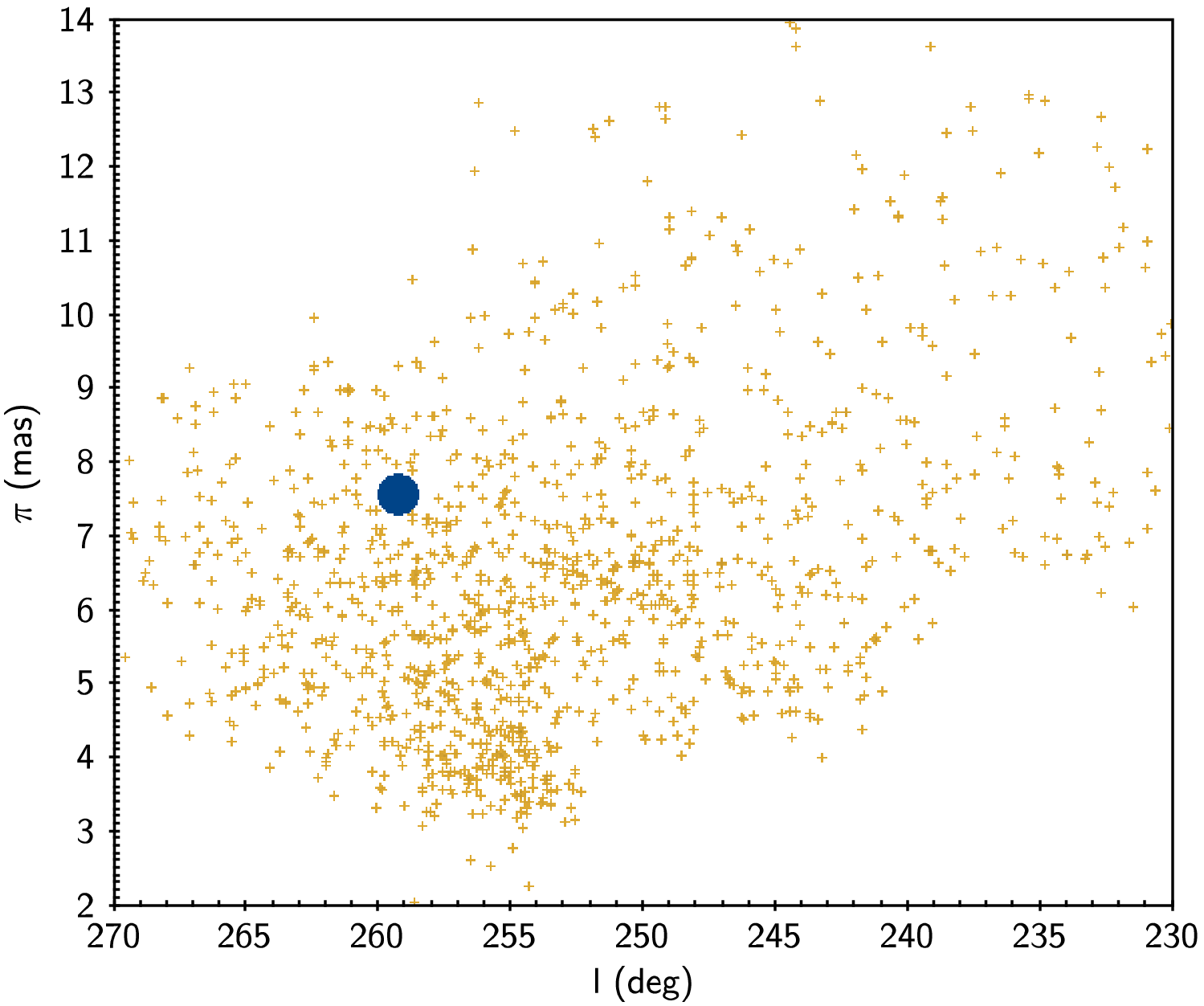}
\plottwo{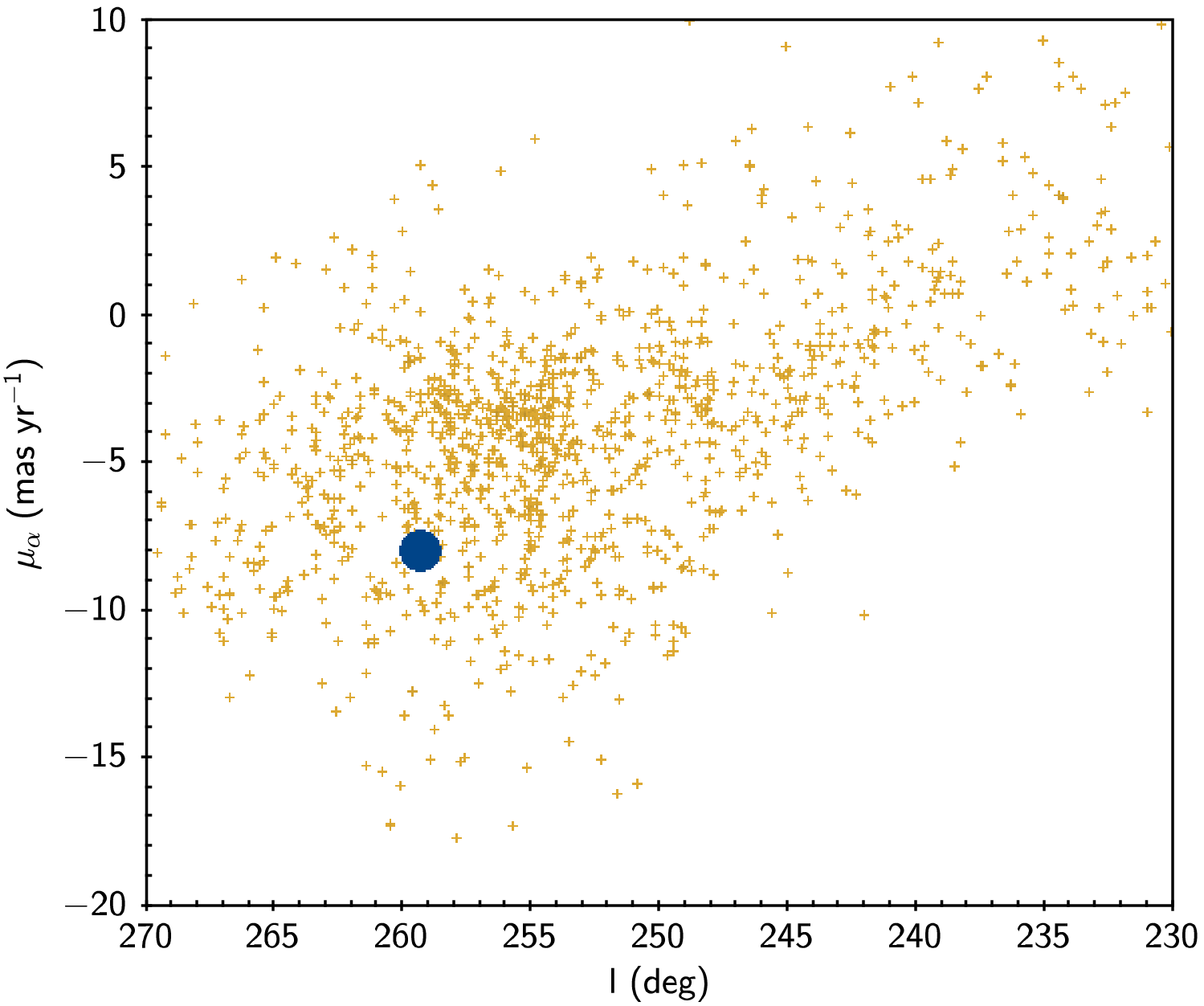}{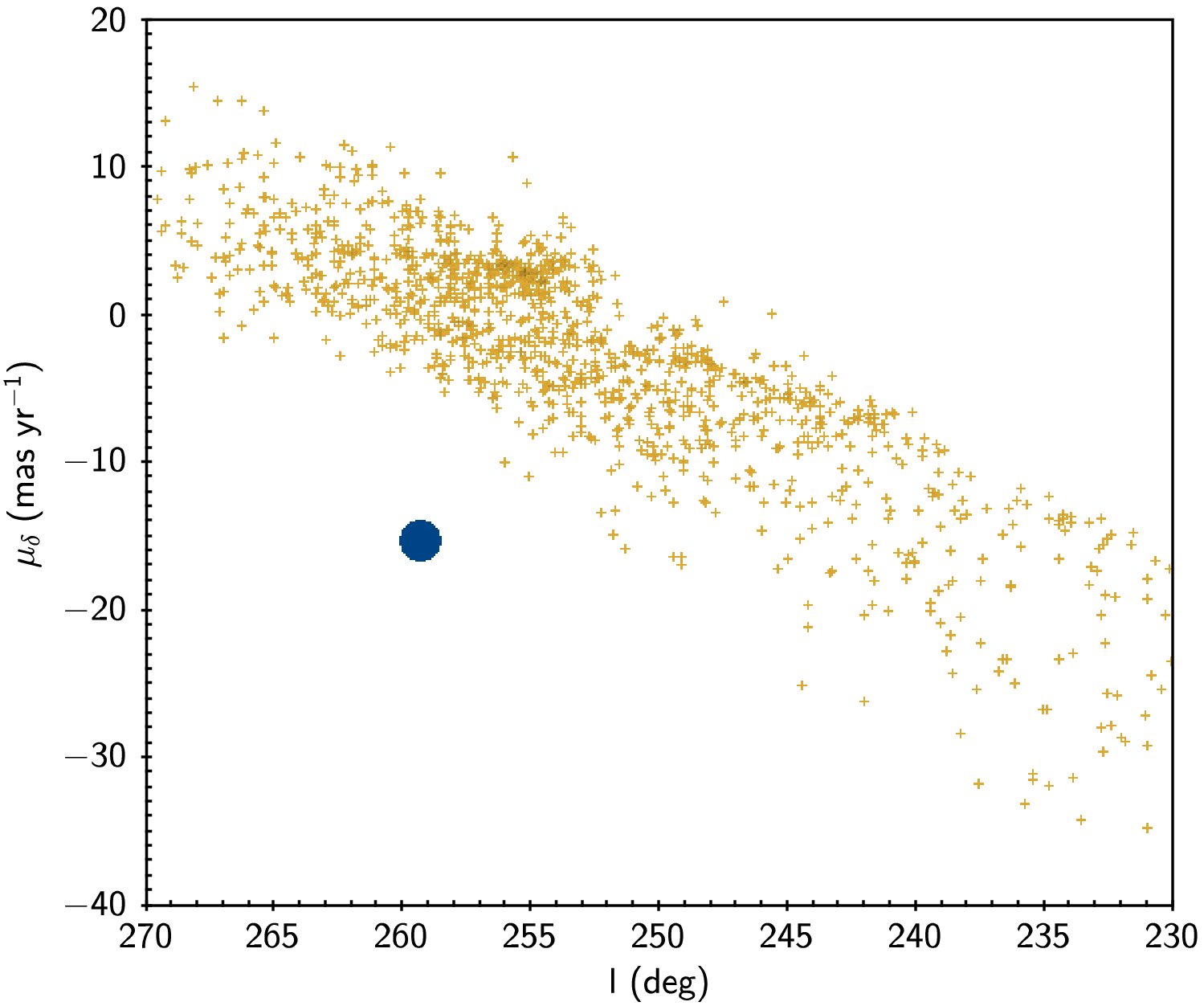}
\caption{Phase space of HD~54236A relative to the phase space of Theia~301.
\label{fig:301}}
\end{figure}

Examining the full phase space correlation between the source and the population as a whole, HD~54236A has good agreement with the 3-dimensional spatial distribution of Theia~301, as well as with $\mu_\alpha$ (Figure \ref{fig:301}). Greater discrepancy is found in $\mu_\delta$, where the source is offset by almost $\sim$10 mas~yr$^{-1}$ (6 km~s$^{-1}$). Such an offset could be due to multiplicity of HD~54236A affecting the astrometric solution, or due to a dynamical evolution of the population over time, gaining sufficient speed over the initial conditions to no longer be recoverable in phase-space clustering.


Recent kinematic analyses have led to new associations being made for co-moving groups of Ursa Major \citep[e.g.,][]{Gagne:2020} and 32~Ori \citep[e.g.,][]{Stauffer:2020}. Similarly, we find that
a possible ancestral link exists between the AB~Dor moving group and Theia~301. Six listed members of the AB~Dor association also have membership to Theia~301 \citep{Messina:2010}. Connection between the stellar censuses of Theia-301 and AB~Dor are tenuous as so few members of AB~Dor are known at distances farther than 80~pc and clustering from \citet{Kounkel:2019} struggles to identify members of moving groups within 100~pc. The observational knowledge gaps in established AB-Dor members beyond 80~pc and clustering within 100~pc converge where we would expect to find shared stellar members between these associations. Even so, there have been 6 stars from \citet{Messina:2010} that are listed as members of AB~Dor are also members of Theia~301, serving as a bridge between the two populations. The two populations are independently estimated to have roughly the same ages. The age of Theia~301 obtained from the population isochrone fitting is 8.29$\pm$0.16 dex, i.e, 134--280~Myr. The age of AB~Dor moving group is estimated at 130--200~Myr \citep{Bell:2015}. Both are consistent with the age of HD~54236 derived through Li I (\S \ref{sub:lithium}) and Ca II (\S \ref{sub:CaHK}) abundances. 

The age of the AB~Dor association's namesake, AB~Dor, has been a subject of debate since the findings of \citet{Close:2005} estimated the age of the lowest-mass component of the system, AB~Dor~C, to be only $\sim$50~Myr. 
%
That same year, 
\citet{Luhman:2005} reported a conservative age range of AB~Dor~C as 75--150 Myr. When the \cite{Close:2005} more youthful age estimate of $\sim$50~Myr is used, the luminosity of AB~Dor~C appears highly discrepant with stellar evolution models predictions. However, when an older age like those suggested by \citet{Luhman:2005} and \citet{Barenfield:2013} is used, much of the luminosity discrepancy disappears.
Assuming that HD~54236 is indeed a member of Theia~301, and further that Theia~301 is linked to the AB~Dor association, our robust age estimates of HD~54236 through lithium abundances (see \S\ref{sub:lithium}) and \ion{Ca}{2} H\&K (\S\ref{sub:CaHK}) would in turn imply an age for AB~Dor that agrees with older age estimates. Our age estimate of $\sim$225~Myr being applied to AB~Dor~C mitigates its luminosity problem. 

\section{Summary and Conclusions}\label{sec:summary}

We have shown that HD~54236 is a hierarchical triple system comprising HD~54236A as a previously unknown EB with stellar masses of 1.18~M$_\odot$ and 1.07~M$_\odot$, and the late-K type star HD~54236B as a wide CPM tertiary, with estimated mass 0.5--0.6~M$_\odot$. We have furthermore shown that the system is young, with a most likely system age of 225$\pm$50~Myr from joint consideration of the lithium abundances measured in the three stars and the strength of the \ion{Ca}{2}~H\&K chromospheric activity index. The system age is determined independently from the lithium abundances and \ion{Ca}{2}~H\&K activity to be 225$\pm$50~Myr. Using a number of photometric observations for tertiary member HD54236~B, \texttt{MINESweeper} and the latest MIST stellar evolutionary models, yield an age estimate of $149^{+173}_{-77}$~Myr for HD~54236B providing additional credence to our reported age estimate. At this age, the solar-mass eclipsing component stars (HD~54236A) are very nearly zero-age main sequence stars, and the wide low-mass tertiary companion (HD~54236B) is near the end of its PMS evolution. All solutions from our analysis are collected in \ref{tab:summary}.

\begin{table}[!ht]
\centering
\caption{HD~54236 System Solutions}\label{tab:summary}
\begin{tabular}{|l|c|}
\tableline
$T_{\rm{eff,1}}$   & $6480\pm 103\ K$ \\ 
$T_{\rm{eff,2}}$   & $6155\pm 155\ K$ \\
$R_{1}$            & $1.128\pm 0.044\ R_\odot$ \\
$R_{2}$            & $1.021\pm 0.044\ R_\odot$ \\
$M_{1}$            &    $1.179\pm0.029$~\msun \\
$M_{2}$            &    $1.074\pm0.027$~\msun \\ 
$i$                & $79.4\pm 0.2^{\circ}$ \\
\tableline
$q$ & $0.9132 \pm 0.012$ \\
$e$ & $0.009 \pm 0.008$ \\
$v_\gamma$ & $-10.0 \pm 0.5$ \kms \\
$a \sin i$ & $9.810 \pm 0.080$ \rsun \\ 
\tableline
Parallax & $7.56\pm0.024$ mas \\ 
$U$ & $9.07\pm0.61$~km~$s^{-1}$\\
$V$ &  $9.07\pm0.61$~km~$s^{-1}$ \\
$W$ & $-4.97\pm0.23$~km~$s^{-1}$ \\
\tableline
\end{tabular}
\end{table}

The HD~54236 system is also interesting from the standpoint of its dynamics, by virtue of representing a tight EB with orbital period of 2.4~d, with a wide tertiary companion that remains unaffected by the "fountain of youth'' which may exist in EBs close enough to enhance their stellar rotation. The wide tertiary companion also represents a statistically valuable object in stellar multiplicity, and the evolution of three-body tidal effects. For example \citet{Tokovinin:2006} finds that, among close binary star systems with orbital periods less than 3~d, the fraction with wide tertiaries is 96\%. This promotes three-body dynamics as the primary mechanism for hardening short-period binaries, such as the short period EB HD~54236A and its tertiary member, HD~54236B. 
Indeed, there may be indications of previous and ongoing dynamics in the HD~54236 system. In particular, our observation of out-of-eclipse variations on a period very similar to but slightly shorter than the 
EB orbital period indicates that at least one component of the eclipsing binary HD~54236A is not rotationally synchronized with the binary orbit, and therefore should still be experiencing tidal effects. Later observations of the EB HD~54236A would be expected to have an out-of-eclipse variation closer to the period as rotational-orbital synchronization is achieved.

The HD~54236 system previously had no known stellar association to which it might belong. When using the $\textrm{BANYAN}~\Sigma$ Multivariate Bayesian Algorithm \citep{Gagne:2018} with the \textit{Gaia} DR2 distance estimates and proper motion information \citep{Bailer-Jones:2018}, HD~54236 was found to have 99\% probability of being a field star. That is, every known (as of 2018) stellar association within 150~pc has a nearly 0\% probability of having HD~54236A among its members. Additionally, its \textit{Gaia} radial velocity estimates were not similar to radial velocity estimates of the systems considered by $\textrm{BANYAN}~\Sigma$. However, the \citet{Kounkel:2019,Kounkel:2020} catalog presented newly discovered of stellar associations, including ``stellar  strings", and found a likely membership to the stellar string Theia~301, which has an estimated kinematic age that is also in very good agreement with our independent age estimate for HD~54236. 

Finally, the AB~Dor association (and thus AB~Dor itself) appears to be linked both kinematically and through a number of shared stellar members to this newly discovered ``stellar  string" Theia~301. Therefore, the age of $\sim$225~Myr that we have found for HD~54236 (and Theia~301) would imply that the AB~Dor system also shares this age. Indeed, applying our age estimate to AB~Dor itself alleviates reported tension between observation and theory that arises for the luminosity of 
AB~Dor~C, when younger age estimates ($\sim$50~Myr) are used. 

\software{C-Munipack (http://c-munipack.sourceforge.net/), ISIS image subtraction package (Alard \& Lupton 1998; Alard 2000; Hartman et al.\ 2004), SExtractor (Bertin \& Arnouts 1996), DAOPHOT II (Stetson, Davis, \& Crabtree 1990), IRAF (Tody 1986, Tody 1993), MINESweeper (Cargyle et al. 2019), MIST (Choi et al. 2019), Lightkurve; s Python package for Kepler and \textit{TESS} data analysis (Lightkurve Collaboration, 2018), astrolibR; used to convert \textit{Gaia} distance, proper motion and radial velocity to galactic coordinates (Chakraborty, Feigelson, \& Babu 2014)}

\acknowledgments
We thank the reviewer for a constructive report and helpful feedback. 
$\textrm{J=MC}^{2}$ would like to thank Erika McIntire, Nicole Megetarian \& S. J. Clark for their continued empowerment of a mother in STEM. K.G.S.\ acknowledges NSF grant AST-1009810. 
We thank A.\ Aarnio for independently checking our
calculation of the $UVW$ space motion of the system.
We also gratefully acknowledge Jaci Cloete and Piet Fourie
for their technical assistance in the operation of the
KELT-South telescope at the South African Astronomical
Observatory.
This research has made use of the SIMBAD database,
operated at CDS, Strasbourg, France. 

\newpage

\bibliographystyle{aasjournal.bst}
\bibliography{biblio.bib}
\end{document}